\documentclass[acmsmall,screen]{acmart}

\AtBeginDocument{%
  }

\newcommand{\tabincell}[2]{\begin{tabular}{@{}#1@{}}#2\end{tabular}}
\usepackage{xspace}
\usepackage{url}

\usepackage[framemethod=tikz]{mdframed}
\newcounter{finding}
\newcommand{\finding}[1]{\refstepcounter{finding}
  \vspace{2.3mm}
 \begin{mdframed}[linecolor=gray,roundcorner=12pt,backgroundcolor=gray!15,linewidth=3pt,innerleftmargin=2pt, leftmargin=0cm,rightmargin=0cm,topline=false,bottomline=false,rightline = false]
  \textbf{Ans. to RQ\arabic{finding}:} #1
 \end{mdframed}
 \vspace{2.3mm}
}
\usepackage{multirow}
\usepackage{listings}
\usepackage{cancel}
\usepackage{enumitem}
\usepackage{threeparttable}

\mdfdefinestyle{listingstyle}{
  linewidth=0.5pt,
  linecolor=black,
  topline=true,
  bottomline=true,
  rightline=false,
  leftline=false,
  innertopmargin=5pt,
  innerbottommargin=5pt,
}

\usepackage{graphicx}
\usepackage{subfigure}

\usepackage{pifont}

\usepackage{listings}
\usepackage{xcolor}
\setcopyright{acmlicensed}
\copyrightyear{2024}
\acmYear{2024}
\acmDOI{XXXXXXX.XXXXXXX}

\begin{document}


\newcommand{\toolName}{\textit{SlsDetector}\xspace}

\title{LLM-Based Misconfiguration Detection for AWS Serverless Computing}


\author{Jinfeng Wen}
\affiliation{%
  \institution{Beijing University of Posts and Telecommunications}
  \city{Beijing}
  \country{China}
}
\email{jinfeng.wen@bupt.edu.cn}

\author{Zhenpeng Chen}
\email{zhenpeng.chen@ntu.edu.sg}
\affiliation{%
  \institution{Nanyang Technological University}
  \city{Singapore}
  \country{Singapore}
}

\author{Federica Sarro}
\email{f.sarro@ucl.ac.uk}
\affiliation{%
  \institution{University College London}
  \city{London}
  \country{United Kingdom}
}

\author{Zixi Zhu}
\email{zhuzx816@163.com}
\affiliation{%
  \institution{Beijing University of Posts and Telecommunications}
  \city{Beijing}
  \country{China}
}

\author{Yi Liu}
\email{liuyi14@pku.edu.cn}
\affiliation{%
  \institution{Advanced Institute of Big Data}
  \city{Beijing}
  \country{China}
}

\author{Haodi Ping}
\email{haodi.ping@bjut.edu.cn}
\affiliation{%
  \institution{Beijing University of Technology}
  \city{Beijing}
  \country{China}
}

\author{Shangguang Wang}
\email{sgwang@bupt.edu.cn}
\affiliation{%
  \institution{Beijing University of Posts and Telecommunications}
  \city{Beijing}
  \country{China}
}

\renewcommand{\shortauthors}{Jinfeng Wen et al.}

\begin{abstract}
Serverless computing is an emerging cloud computing paradigm that enables developers to build applications at the function level, known as serverless applications. Amazon Web Services (AWS), the leading provider in this domain, provides the Serverless Application Model (AWS SAM), the most widely adopted configuration schema for configuring and managing serverless applications through a specified file. However, misconfigurations pose a significant challenge in serverless development.
Traditional data-driven techniques, which learn configuration patterns from historical data to identify anomalies, may struggle with serverless applications because the complexity of serverless configurations hinders pattern recognition, and it is challenging to gather complete datasets that cover all possible configurations.
% Traditional data-driven techniques, which learn configuration patterns from datasets to identify anomalies, struggle with the limitations of incomplete or incorrect datasets and the structural complexity of serverless application configurations.
% Traditional data-driven techniques, which learn configuration patterns from datasets to identify anomalies, are not well-suited to practical use in serverless computing. These approaches face limitations, such as incomplete or incorrect datasets, and they struggle with the structural complexity of configurations of serverless applications. 
Recent advancements in Large Language Models (LLMs) have shown promise in tackling various software engineering tasks. Leveraging vast amounts of publicly available data during pre-training, LLMs can have the potential to assist in identifying and explaining misconfigurations in serverless applications.
% cannot be directly adopted in practice due to the inherent limitations of these methods (e.g., incomplete or incorrect dataset) and the complexity of serverless configuration features with intricate structures. 
% Recent work has shown the large potential of Large Language Models (LLMs) in various software engineering tasks. By being pre-training on vast amounts of publicly available data, the models are capable of find and explain configuration problems of serverless applications.

In this paper, we introduce \toolName, the first framework leveraging LLMs to detect misconfigurations in serverless applications. \toolName utilizes effective prompt engineering with zero-shot learning to identify configuration issues. It designs multi-dimensional constraints specifically tailored to the configuration characteristics of serverless applications and leverages the Chain of Thought technique to enhance LLMs inferences, alongside generating customized structured responses.
We evaluate \toolName on a curated dataset of 110 configuration files, which includes correct configurations, real-world misconfigurations, and intentionally injected errors.
Our results show that \toolName, based on ChatGPT-4o (one of the most representative LLMs), achieves a precision of 72.88\%, recall of 88.18\%, and F1-score of 79.75\%, outperforming state-of-the-art data-driven approaches by 53.82, 17.40, and 49.72 percentage points, respectively. 
Furthermore, we investigate the generalization capability of \toolName by applying recent LLMs, including Llama 3.1 (405B) Instruct Turbo and Gemini 1.5 Pro, with results showing consistently high effectiveness across these models.

\end{abstract}

\ccsdesc[500]{Software and its engineering~Software configuration management and version control systems}
\ccsdesc[500]{Computer systems organization~Cloud computing}

\maketitle

\section{Introduction}\label{sec:introduction}

Serverless computing is an emerging cloud computing paradigm that allows developers to build and run applications, known as \textit{serverless applications}, without managing underlying infrastructure tasks~\cite{wen2023literature}. It has been widely adopted across diverse application domains~\cite{yu2021gillis,shankar2020serverless,ao2018sprocket}, attracting growing interest from research communities, such as Software Engineering (SE)~\cite{wen2023literature} and Systems~\cite{LiGCCHG22}, and from industry.
To support the development and execution of serverless applications, leading cloud providers have introduced serverless platforms. Among these providers, Amazon Web Services (AWS) stands out as the leader in serverless computing~\cite{wen2023literature, JonasCoRR2019, Wen21challenges}.

Serverless computing supports two primary service models: Function-as-a-Service (FaaS) and Backend-as-a-Service (BaaS)~\cite{hassan2021survey, eismann2021state}. FaaS allows developers to build applications as small, event-driven functions (i.e., serverless functions), while BaaS provides ready-to-use cloud services such as storage (e.g., AWS S3~\cite{S3}), database, and API gateway management. This collaboration between FaaS and BaaS enables developers to efficiently create serverless applications.

To configure and manage functions and required cloud resources for serverless applications, AWS provides the Serverless Application Model (AWS SAM)~\cite{sam}, the most widely adopted configuration schema in the serverless computing practice~\cite{popular1sam, examplesam, SAR}. It can streamline the development process and reduce complexities associated with resource management in serverless applications. 
% These misconfigurations lead to issues such as crashes, incorrect triggers, and deployment errors~\cite{Wen21challenges}. 

However, misconfigurations have emerged as a major challenge in serverless application development~\cite{sun2020testing, li2018confvd, Wen21challenges}. 
Misconfigurations in serverless computing can lead to significant security vulnerabilities and operational issues. For example, as reported~\cite{realexamplemisconfiguration1}, a coronavirus testing company exposed over 50,000 patients' scanned IDs and thousands of COVID-19 test results due to a misconfiguration in an AWS S3 bucket, used in conjunction with serverless applications. Similarly, another company experienced a data breach affecting 4.9 million customers due to API misconfigurations within serverless environments~\cite{realexamplemisconfiguration2}.
These examples underscore that misconfigurations are not isolated incidents but represent systemic issues that pose significant risks to serverless applications, suggesting the urgent need for effective misconfiguration detection for serverless computing.

% the urgent need for effective misconfiguration detection for serverless computing.

% Addressing these issues demands substantial expertise and is often time-consuming and labor-intensive, making automated misconfiguration detection a crucial topic of research.

% Considerable expertise and effort are required to identify the root cause of the problem. Thus, it is crucial to provide an automated and effective tool to detect misconfigurations in serverless applications~\cite{wen2022software}.

% However, misconfigurations are a major cause of software failures~\cite{sun2020testing, li2018confvd} and are prevalent in serverless application development~\cite{TEST21, TEST3, TEST14, TEST40, TEST58}. They are the well-known major challenge that developers face~\cite{Wen21challenges}, arising from configurations for serverless functions, resources, and customization requirements. These issues can potentially result in application failures, unexpected deployments, and incorrect event triggers~\cite{Wen21challenges}. 
% This often leaves developers without a clear solution, forcing them to seek technical support. When support engineers encounter ambiguous symptoms, they have to reproduce the reported configurations to diagnose potential problems, a process that can be time-consuming.
% Developers may resort to guesswork or unreliable sources for configuring serverless applications, wasting time, effort, and computing resources. 
% Therefore, it is crucial and highly desirable to provide an effective misconfiguration detection tool for serverless applications~\cite{wen2022software}.

Misconfigurations have become one of the major causes of system software failures~\cite{xu2015systems}. 
Despite the promise of existing data-driven methods for misconfiguration detection in other scenarios~\cite{zhang2014encore, santolucito2017synthesizing, zhou2023drive, santolucito2016probabilistic}, they have low effectiveness to serverless computing. Data-driven approaches, which rely on anomaly detection or pattern recognition based on training data, suffer from limitations such as incomplete or incorrect datasets~\cite{zhang2014encore, zhou2023drive}. Additional strategies that incorporate extensive knowledge, such as predefined templates and official documentation, lack flexibility and adaptability. These problems make the data-driven approach not enough to detect configuration problems of serverless applications. Moreover, serverless application configurations involve intricate structures, including domain-specific languages, complex dependency relationships, and nested objects across over 800 cloud resource types, which further complicates their detection.

% Unfortunately, existing approaches for detecting misconfigurations applied in other domains are not applicable to serverless applications. Data-driven methods have emerged as promising and widely adopted techniques, commonly identifying potential errors as anomalies or deviations from normal patterns based on training data of example configurations~\cite{zhang2014encore, santolucito2017synthesizing, zhou2023drive, santolucito2016probabilistic}. However, the approach's effectiveness heavily relies on the completeness and correctness of the training data, which poses an inherent limitation in ensuring comprehensive patterns. Some strategies~\cite{} incorporating extensive knowledge (e.g., official documentation, predefined templates, and syntax) derive from manual efforts into their design. This causes additional problems with flexibility and generalizability, finally, making them unsuitable for direct use to serverless applications with varying configuration syntax. Serverless application configurations often involve more complex structures, including not only specialized domain-specific languages and key dependency relationships but also nested objects of cloud resource types (more than 800 resource categories), further complicating their adaptation to existing approaches.

Recent advancements in Large Language Models (LLMs) offer a promising new approach to this problem. LLMs have demonstrated significant success in various SE tasks, such as code summarization~\cite{ahmed2022few}, program repair~\cite{fan2023automated}, unit test generation~\cite{yuan2024evaluating}, and log parsing~\cite{xu2024divlog}. Trained on vast amounts of publicly available data, LLMs can potentially understand and recognize configuration patterns, best practices, and common pitfalls in serverless application configurations. This makes LLMs well-suited for detecting potential misconfigurations in serverless applications.

% The recent advent and success of Large Language Models (LLMs) in performing complex software tasks~\cite{}, such as XX, XX, and XX. LLMs are trained on vast amounts of publicly available data, which likely includes serverless application configuration fragments, documentation, relevant articles, and discussions from Q\&A websites. This extensive training enables LLMs to recognize patterns, best practices, and common pitfalls and give the corresponding explanation, making them well-suited for efficiently finding configuration issues in serverless applications. This suggests a promising avenue for detecting potential misconfigurations of serverless applications. 

% To address the gap, we propose the first LLM-based misconfiguration detection approach for serverless applications, \toolName. 
% This approach avoids the inherent limitations of existing approaches. 
% The core of \toolName leverages advanced LLMs to perform complex detection tasks. 
% The recent advent and success of LLMs in performing complex software tasks~\cite{}, suggests a promising avenue for detecting potential misconfigurations of serverless applications. 
% LLMs are trained on vast amounts of publicly available data, which likely includes serverless application configuration fragments, documentation, relevant articles, and discussions from Q\&A websites. This extensive training enables LLMs to recognize patterns, best practices, and common pitfalls and give the corresponding explanation, making them well-suited for efficiently finding configuration issues in serverless applications.

In this paper, we introduce \toolName, the first LLM-based framework specifically designed to detect misconfigurations in serverless applications. 
By leveraging advanced prompt engineering in conjunction with zero-shot learning, which requires no prior examples, \toolName efficiently identifies configuration problems with minimal effort.
This marks a significant advancement in detecting misconfigurations within serverless environments.
% marking a significant advancement in the automation of serverless misconfiguration detection. 
\toolName accepts a configuration file of the serverless application as input and outputs detected misconfigurations along with detailed explanations for each issue. 
% \toolName takes a configuration file of the serverless application as the input, and output detected misconfigurations along with reasons that explain them.
% Specifically, by combining advanced prompt engineering with zero-shot learning, \toolName efficiently identifies configuration issues with minimal effort, representing a substantial advancement in automated serverless configuration detection.
% \toolName leverages advanced prompt engineering with zero-shot learning to identify configuration issues with minimal input, making a significant leap in serverless configuration detection. 
To achieve this, \toolName features a prompt generation component that dynamically integrates the configuration file, task description, multi-dimensional constraints, and customized responses. Multi-dimensional constraints are tailored to the specific characteristics of serverless applications, taking into account resource types, configuration entries, values, and different levels of dependencies. This context-aware design provides targeted guidance for detection. Furthermore, \toolName employs the Chain of Thought technique~\cite{CoT, chu2023survey}, a reasoning strategy that enhances the problem-solving process, into these constraints. 
The customized response provides the content demand and format demand of LLM outputs, ensuring that responses are not only structured but also actionable answers aligned with detailed explanations.

% In this paper, we propose the first LLM-based misconfiguration detection approach for serverless applications, \toolName. To the best of our knowledge, \toolName is also the first automated misconfiguration detection tool for serverless applications. Specifically, \toolName leverages the effective prompt engineering with zero-shot learning. It introduces a goal generation module and an answer requirement module. The goal generation module contains the content of the target configuration file, a directive question, and novel multi-dimensional constraints specifically for configuration characteristics of serverless applications. The answer requirement module involves customizing response content and format to obtain the desired answers.

To evaluate \toolName, we curate a dataset of 110 configuration files, including 26 correctly configured files, 58 with real-world misconfigurations, and 26 with injected errors. Our results show that \toolName, based on ChatGPT-4o (one of the most representative LLMs known for outstanding performance), achieves a precision of 72.88\%, recall of 88.18\%, and F1-score of 79.75\%. It outperforms the state-of-the-art data-driven approach by 53.82, 17.40, and 49.72 percentage points in precision, recall, and F1-score, respectively. 
We further explore the generalization capability of \toolName on other representative LLMs, including Llama 3.1 (405B) Instruct Turbo and Gemini 1.5 Pro. The results show that \toolName consistently achieves high effectiveness across these models.

In summary, this paper makes the following contributions:
 \begin{itemize}[leftmargin=*]
    \item We present \toolName, the first LLM-based approach specifically designed for detecting misconfigurations in serverless computing. 
    % an open-source evaluation dataset in a replication package~\cite{ourdata}, along with the code, data, and results from this paper, to support future research and replication.
    
    \item We conduct an empirical study using our benchmark dataset to evaluate the effectiveness of our misconfiguration detection approach, demonstrating that it outperforms baseline methods.
    % We conduct an empirical study on the effectiveness of misconfiguration detection techniques in serverless computing using our dataset, demonstrating it outperforms baseline methods.

     % construct an open-source evaluation dataset of serverless application configuration files, accompanied by ground truth results to support evaluation purposes. release...

    % \item We offer a replication package~\cite{ourdata} containing the code, dataset, and evaluation results used in this paper to facilitate future research and replication.
    % \wjf{We,  Leveraging this dataset, empirical study,...}

\end{itemize}

\section{Background}\label{sec:background}

% We provide an overview of serverless computing and serverless application configurations, along with a real-world example to illustrate specific configuration details.

\subsection{Serverless Computing}

% Serverless computing is an emerging cloud computing paradigm that allows developers to focus on application logic without managing infrastructure tasks. 

% In serverless computing, applications that developers develop are called \textit{serverless applications}. They are implemented as event-driven serverless functions and related cloud services. Specifically, serverless functions are the core of serverless computing, representing the implementation of business logic, written in popular programming languages~\cite{eismann2021state, wen2023literature}. Cloud services are provided by the corresponding cloud providers to simplify the implementation of backend complex functionalities and integrate them into serverless functions.

Applications developed within the serverless computing paradigm are referred to as serverless applications.
% In serverless computing, applications \wjf{developed by developers, based on paradigm} are referred to as serverless applications. 
These applications are built around event-driven serverless functions, which represent the core business logic. They are complemented by associated cloud services that facilitate the integration of backend functionalities. This combination streamlines the development process~\cite{eismann2021state, wen2023literature}.
% and associated cloud services. Serverless functions, which form the backbone of serverless computing, are used to represent application business logic and are written in different languages~\cite{eismann2021state, wen2023literature}. Complementing these functions, cloud services provided by cloud providers facilitate the integration of complex backend functionalities, streamlining the development process. 
During the development and deployment of serverless applications, developers define essential execution settings. These settings include the runtime environment, memory allocation, timeout duration, predefined event triggers, and required cloud resources for the serverless applications.

\subsection{Serverless Application Configurations: AWS SAM}

% Developers can utilize a specified configuration file (e.g., YAML file) to define the execution settings of serverless applications. Generally, relationships between serverless functions and pre-defined events are not explicitly specified in user code, as serverless functions are event-driven. However, they can be described in this configuration file in a brief way to automatically handle infrastructure provisioning. Thus, such a specific configuration file is crucial for the development process of serverless applications.

Developers leverage specified configuration files, such as YAML files, to define the execution settings of serverless applications. In serverless computing, serverless functions are inherently event-driven, meaning that the relationships between functions and predefined events are not explicitly detailed in the application code. Instead, these relationships are succinctly captured in the configuration file, which automates infrastructure provisioning. Thus, the configuration file plays a crucial role in the development process of serverless applications.

% serverless architectures, relationships between functions and pre-defined events are not explicitly specified in the application code, as serverless functions are inherently event-driven. Instead, these relationships are succinctly described in the configuration file, which automates infrastructure provisioning. As a result, the configuration file plays a crucial role in the serverless application development process.

Among mainstream serverless platforms, AWS Lambda employs a widely used configuration schema~\cite{popular1sam, examplesam} known as the AWS Serverless Application Model (AWS SAM)~\cite{sam}. AWS SAM enables developers to easily reuse proven configurations, streamlining the development and deployment of serverless applications. In contrast, other platforms, such as Google Cloud Functions~\cite{google} and Microsoft Azure Functions~\cite{azure}, lack a formal configuration schema. They rely on command-line interfaces or platform consoles to manually manage key settings and required resources. This manual way lacks standardization and the availability of configuration datasets for analysis. \textbf{Given AWS Lambda's widespread use and the advantages offered by AWS SAM's configuration schema, our paper focuses on analyzing the configurations of serverless applications built using AWS SAM.}

AWS SAM uses a YAML-based configuration file format with specialized template specifications. It builds upon and extends AWS CloudFormation~\cite{CloudFormation}, which is primarily used for provisioning and configuring non-serverless cloud resources. AWS SAM introduces a syntax specifically designed for defining and managing both serverless infrastructure (spanning nine categories~\cite{serverlessresource}) and non-serverless infrastructure (covering over 800 categories~\cite{awsresource}).

% AWS SAM employs a YAML-based configuration file format with specialized template specifications. Building on and extending AWS CloudFormation~\cite{CloudFormation}, which is primarily used to provision and configure non-serverless cloud resources, AWS SAM introduces a syntax specifically designed for defining and managing both serverless cloud infrastructure (spanning 9 categories~\cite{serverlessresource}) and non-serverless cloud infrastructure (covering over 800 categories~\cite{awsresource}).

Serverless application configurations are complex and exhibit unique characteristics. Unlike the simple ``flat'' key-value pair format commonly seen in prior configuration studies~\cite{santolucito2017synthesizing, chen2020understanding, wang2023understanding, sun2020testing, xu2013not}, serverless application configurations feature intricate structures, including objects, lists, maps, and nested elements. Each cloud resource type is represented by custom-named objects, which contain specific configuration entries and their corresponding values. These values can be strings, lists, maps, or even nested objects representing other cloud resources.
Additionally, serverless configurations introduce resource types specific to serverless environments (e.g., ``AWS::Serverless::Function'') and attributes unique to serverless applications (e.g., \texttt{Handler}, \texttt{MemorySize}, \texttt{Timeout}). This exhibits that AWS SAM YAML files function as domain-specific languages within the serverless computing domain, increasing the complexity of configurations.

\subsection{An Example of the Configuration File} \label{sec:example}

\begin{figure}[t]
\tiny
\vspace{-3mm}
\begin{mdframed}[style=listingstyle]
\begin{lstlisting}[breaklines=true, numbers=left, numbersep=5pt, numberstyle=\tiny\hfill]
AWSTemplateFormatVersion: '2010-09-09'
Transform: AWS::Serverless-2016-10-31
Description: Lambda that responds to S3 events
Parameters:
    PreExistingBucket:
        Description: "Does an existing bucket exist (not managed by cloudformation)"
        Type: String
        Default: 'no'
        AllowedValues:
            - 'yes'
            - 'no'
        ConstraintDescription: must specify yes or no.
Conditions:
    NeedsSomeBucket: !Equals [!Ref PreExistingBucket, 'no']
Resources:
    BucketEventConsumer:
        Type: AWS::Serverless::Function
        Properties:
            Handler: BucketEventConsumer.main.lambda_handler
            Runtime: python3.6
            CodeUri: bundle.zip
            Events:
                CreateMetaEvent:
                    # Condition: NeedsSomeBucket
                    Type: S3
                    Properties:
                        Bucket: !Ref SomeBucket
                        Events: "s3:ObjectCreated:*"
                        Filter:
                            S3Key:
                                Rules:
                                    - Name: suffix
                                      Value: meta.json
    SomeBucket:
        Condition: NeedsSomeBucket
        Type: AWS::S3::Bucket
        Properties:
            BucketName: 'some-bucket-somewhere'
        DeletionPolicy: Retain
\end{lstlisting}
\end{mdframed}
\caption{An configuration file example of serverless applications.}
\label{fig:configuration-example}
\vspace{-3mm}
\end{figure}

We provide a real-world configuration file example~\cite{TEST21} from GitHub, a widely used platform for studying developer issues~\cite{kourtzanidis2020reposkillminer, kavaler2017perceived}, as shown in Fig.~\ref{fig:configuration-example}. In this example, the developer created a serverless function that responds to events from the AWS S3 storage service~\cite{S3}. However, this configuration failed during deployment. Resolving this issue required nearly 20 rounds of communication involving 26 people and spanned almost five years before a correct solution was found.
The root cause was the unsupported \texttt{Condition} entry mistakenly added on line 24. This example underscores the critical need for an effective approach to detect misconfigurations in serverless applications early. Such an approach would quickly pinpoint potential issues, reducing the time, effort, and communication overhead required to troubleshoot and resolve misconfigurations.

% , ultimately streamlining the serverless application development process.

% We provide a real-world configuration example~\cite{TEST21} about serverless applications from GitHub, a commonly used platform for studying developers' issues, as shown in Figure~\ref{fig:configuration-example}. In this example, the developer defined a serverless function that responds to events from the storage service, AWS S3~\cite{S3}. However, this configuration file failed when deployed. To address it, nearly 20 rounds of communication involving 26 persons took place. Moreover, it took almost 5 years to resolve and provide a correct solution. The reason is that the \texttt{Condition} configuration entry of line 24 is not supported, but the developer mistakenly added it. This underscores the importance of designing an effective approach for detecting misconfigurations in serverless applications. Such an approach can quickly identify potential misconfigurations, significantly streamlining the development process and reducing the time and effort required to troubleshoot configuration issues.

We explain this configuration file example. The content is mainly structured into two sections: \texttt{Transform} (line 2) and \texttt{Resources} (lines 15-39). The \texttt{Transform} section identifies the file as an AWS SAM template with the value ``AWS::Serverless-2016-10-31.'' 
The \texttt{Resources} section defines the required execution settings through \textbf{resource types}. The ``AWS::Serverless::Function'' resource type (named ``BucketEventConsumer'') aims to configure a serverless function, while ``AWS::S3::Bucket'' (named ``SomeBucket'') represents an AWS S3 bucket, a non-serverless resource that frequently interacts with serverless functions. The ``BucketEventConsumer'' object includes \textbf{configuration entries} such as the handler function (line 19), runtime environment (line 20), code location (line 21), and a predefined event (lines 22-33). These entries are allocated \textbf{values} that conform to the constraints. For example, \texttt{Runtime} is set to ``python3.6'' (line 20). 
In this example, the event source is set to S3 (line 25) using a nested object. The function is triggered when an S3 object is created (lines 27-28) that meets the filter rule specified as key-value pairs (lines 31-33). Specifically, the function is invoked when a file in the ``SomeBucket'' S3 bucket ends with ``meta.json'' (lines 32-33). \texttt{Name} from line 32 and \texttt{Value} from line 33 need to appear together, indicating \textbf{entry dependencies}. 
Line 27 illustrates a relationship between \texttt{Bucket} value and the ``AWS::S3::Bucket'' resource in line 34, showing that the \textbf{value dependencies} of one configuration value depend on other values.

In addition to the core sections, other parts of the configuration file are also important. The \texttt{AWSTemplateFormatVersion} section (line 1) specifies the template's capabilities, with the current valid format version being ``2010-09-09''~\cite{FormatVersion}. 
% This entry is generally the first in the template. 
The \texttt{Description} section (line 3) provides a textual description of the template. The \texttt{Parameters} section (lines 4-12) defines values that are passed to the template at runtime. The ``PreExistingBucket'' parameter accepts either ``yes'' or ``no'' as values. The \texttt{Conditions} section (lines 13-14) controls resource creation or property assignment based on the value of a parameter. The ``NeedsSomeBucket'' condition checks if the ``PreExistingBucket'' parameter is set to ``no''. If true, the condition evaluates to true, otherwise, it evaluates to false.

\subsection{LLMs for SE}

% The application of LLMs to downstream tasks has become a pivotal area of research. Some methods related to LLMs often involve fine-tuning the model by updating its parameters based on specific downstream datasets. While this kind of method can yield high task-specific performance, it generally requires substantial computational resources and access to high-quality data, which can limit its practicality in certain contexts. 

The application of LLMs to downstream tasks has become a significant area of research in SE~\cite{ahmed2022few, fan2023automated, yuan2024evaluating, xu2024divlog}. Recent studies~\cite{lian2023configuration, yuan2024evaluating} have demonstrated the potential of prompt engineering to achieve impressive performance across various tasks. 
Prompt engineering offers a flexible and resource-efficient way to utilize LLMs by adapting models to specific task requirements through carefully designed prompts. 
% These prompts help guide LLMs in extracting relevant task-specific knowledge by identifying input-output patterns pertinent to the task.
% Prompt engineering provides a more flexible and resource-efficient approach to leveraging LLMs for downstream tasks. It adapts models to task requirements through carefully designed prompts. These prompts guide LLMs in extracting task-specific knowledge by identifying input-output patterns relevant to the task. 
In contrast, the fine-tuning method involves updating the model's parameters using specific downstream datasets. However, it often requires substantial computational resources and access to high-quality data, limiting its practicality in other contexts.

This paper aims to capitalize on the strengths of prompt engineering by developing specialized prompts to detect misconfigurations in serverless applications. This ensures effective detection while avoiding the computation and data demands associated with fine-tuning methods.

% This paper seeks to harness the strengths of prompt engineering by designing tailored prompts that detect misconfigurations in serverless applications, ensuring detection effectiveness without the computational overhead of fine-tuning approaches.

\section{Our approach: \toolName}\label{sec:tool}

% We introduce \toolName, an LLM-powered framework designed for detecting misconfigurations in serverless applications. \toolName processes a configuration file as input and generates structured answers, which include a list of detected misconfigurations alongside explanations for each error. 
% The framework can flexibly support different mainstream LLMs, including GPT, CodeLlama, and Gemini. Next, we introduce the overview and key components of \toolName.

We present \toolName, an LLM-based framework designed to detect misconfigurations in serverless applications. \toolName takes a configuration file of the serverless application to be detected as input and outputs structured results, providing a list of detected misconfigurations along with detailed explanations for each issue. The framework is designed to be adaptable, supporting various LLMs. In the following sections, we provide an overview of \toolName and outline its component.

\subsection{Overview}

\begin{figure*}[t]
	\centering
    \includegraphics[width=0.92\textwidth]{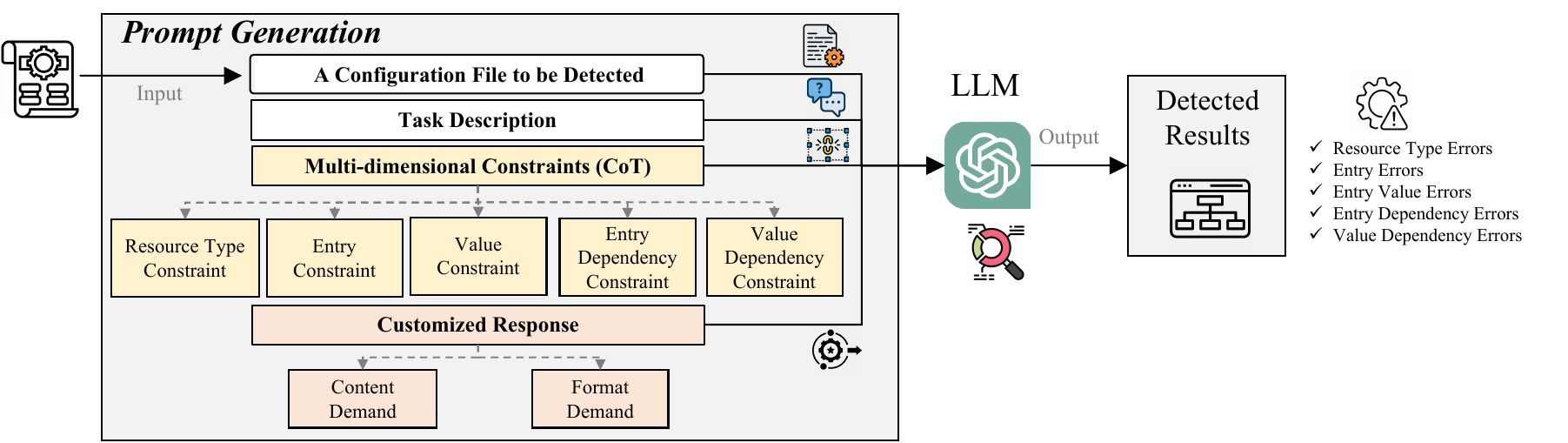}
    \vspace{-3mm}
    \caption{The overview of our approach \toolName.}
    % \vspace{-1mm}
    \label{fig:overview}
    \vspace{-3mm}
\end{figure*}

Fig.~\ref{fig:overview} shows an overview of \toolName. It converts a misconfiguration detection request into a meticulously constructed prompt for LLMs. We employ zero-shot learning to minimize reliance on external sample configurations. This technique, which requires no prior examples, is a popular optimization technique~\cite{yuan2024evaluating, imran2024uncovering, xie2023chatunitest}. While many studies~\cite{lian2023configuration, yin2024multitask, xu2024divlog} have utilized few-shot learning to improve effectiveness by learning from examples during inference, it relies heavily on the quality and selection of labeled samples. In contrast, zero-shot learning avoids the cost and effort associated with sample collection and curation, making it the preferred technique for our framework.
% the structural complexity and unique nature of serverless application configurations make it difficult to collect \wjf{a comprehensive set of examples} that cover all configuration types. 

In \toolName, we design a prompt generation component to construct a tailored prompt focused on the objective of detecting misconfiguration in serverless applications. This prompt is structured into four parts, where multi-dimensional constraints are core of \toolName and highly context-aware, shown in Fig.~\ref{fig:overview}.
% : (i) the inputted configuration file, (ii) a directive task description for LLMs, and (iii) multi-dimensional constraints, and (iv) the customized response. The core of \toolName is to design multi-dimensional constraints, which are highly context-aware and based on characteristics of serverless configurations.
% consists of two core components for constructing tailored prompts in serverless computing: the \textit{Purpose Generation} module and the \textit{Answer Requirement} module. The \textit{Purpose Generation} module captures the main objective of the query, while the \textit{Answer Requirement} module customizes the desired format and details of the response. 
Once the prompt is constructed, it is sent to the LLM, which generates the final output. 
% Note that \toolName does not depend on labeled configuration samples when querying LLMs, thereby eliminating the cost and effort required for data collection and curation.
Next, we introduce the prompt generation component in detail.

% In this situation, \toolName has two key components to construct the tailored prompt in serverless computing: a \textit{Purpose Generation} module and a \textit{Answer Requirement} module. The \textit{Purpose Generation} module is used to reflect the core purpose of the query, while the \textit{Answer Requirement} module is used to customize the desired response. After constructing the prompt, it is sent the query to the LLM and generates the final output.
% Note that \toolName does not rely on labeled configuration samples in querying LLMs, eliminating the cost and effort associated with collecting and curating such data. 

 % (i) the target configuration file, (ii) a detailed description of the query, (iii) multi-dimensional constraints based on serverless configuration characteristics, and (iv) customized requirements for the format and content of the response.

\subsection{Prompt Generation}

\begin{figure*}[t]
\centering
    \includegraphics[width=0.9\textwidth]{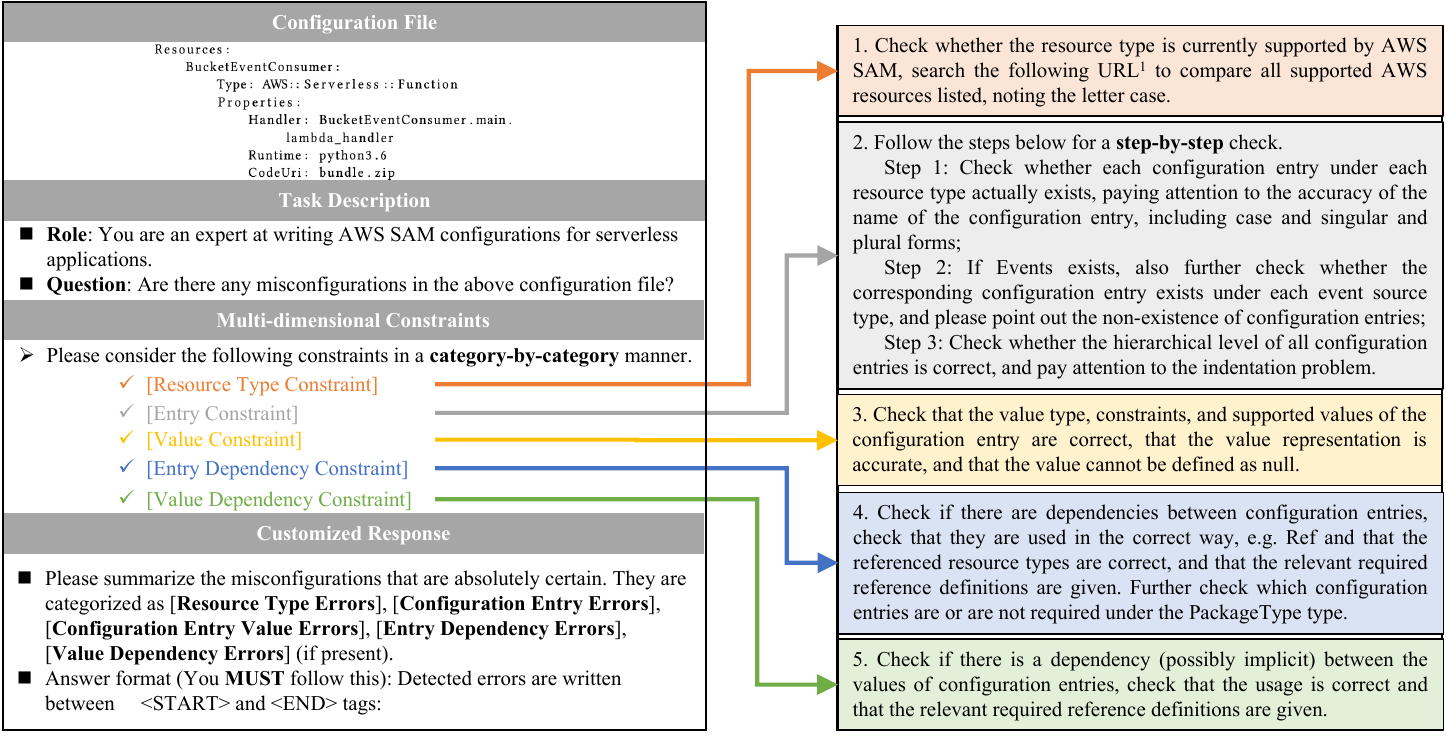}
    \vspace{-3mm}
    \caption{The prompt structure of \toolName.}
    % \vspace{-1mm}
    \label{fig:goalprompt}
    \vspace{-3mm}
\end{figure*}

% \begin{lstlisting}[caption={The prompt of LCoT method.}, label={lst:LCoTprompt}]
% [A Configuration File to be Detected]

% Transform: AWS::Serverless-2016-10-31
% ...

% [Queastion Description] Are there any misconfigurations in the above configuration file? 
% [Multi-dimensional Constraints] Please consider the following constraints in a category-by-category manner.
%     (1) (Type Constraint); 
%     (2) (Entry Constraint);
%     (3) (Value Constraint); 
%     (4) (Entry Dependency); 
%     (5) (Value Dependency).
% \end{lstlisting}

% \subsubsection{Prompt Structure} 

% In this section, we design the partial prompt related to the core objective of the misconfiguration detection query in serverless computing. It mainly consists of three parts: (i) the target configuration file, (ii) a directive question for LLMs, and (iii) multi-dimensional constraints based on the characteristics of the serverless configuration. Figure~\ref{fig:goalprompt} illustrates an example of this partial prompt generated by the purpose generation module in \toolName.

% We introduce our prompt content generated by the prompt generation component, including (i) the configuration file to be detected, (ii) a directive task description for LLMs, (iii) elaborate multi-dimensional constraints, and (iv) customized response.
% Fig.~\ref{fig:goalprompt} provides our prompt structure.

We present the prompt content generated by the prompt generation component, which includes: (i) the configuration file to be analyzed, (ii) a task description for the LLMs, (iii) detailed multi-dimensional constraints, and (iv) a customized response. Fig.~\ref{fig:goalprompt} illustrates our prompt structure.

% in \toolName. 

% We will explain these parts.

% we design a partial prompt focused on the core objective of detecting misconfigurations in serverless computing. The prompt is structured into three key parts: (i) the target configuration file, (ii) a directive question for LLMs, and (iii) multi-dimensional constraints based on the unique characteristics of serverless configurations. Figure~\ref{fig:goalprompt} provides this partial prompt, generated by the purpose generation module in \toolName.

% Configuration file

% Task description

% constraints

% \subsubsection{Target Configuration File}

% To detect the target configurations, \toolName reads the content of the configuration files. 
% While serverless application configuration files are YAML-based and can be processed as key-value pairs using standard tools, we choose a \wjf{text-based reading approach}. 
% This ensures that indentation between configurations is preserved, which may be critical for certain validation or analysis tasks.

% The content of the target configuration file needs to be read to be checked by \toolName. Although the configuration files of serverless applications are YAML-based files, which have specified processing way for this kind of file type in a key-value format, we still use text-read way to obtain their configuration content. This can preserve indentation information between configurations.

\subsubsection{Task Description} 

The task description includes the following elements: (i) a role-playing instruction designed to enhance the LLM’s ability to detect misconfigurations, which is a common prompt optimization technique~\cite{yuan2024evaluating, dong2024self}; and (ii) a task description instruction. In our scenario, the role is designed as ``You are an expert at writing AWS SAM configurations for serverless applications'', while the task description asks, ``Are there any misconfigurations in the above configuration file?''. These elements are carefully crafted to clearly outline the tasks the LLM needs to complete within the assigned role.

% ``You are an expert at writing AWS SAM configurations for serverless applications'', while the question is phrased as: ``Are there any misconfigurations in the above configuration file?'' This question is crafted to clearly define the tasks the LLM needs to accomplish within the specified role.

\subsubsection{Multi-dimensional Constraints}

% We employ a zero-shot learning approach, which requires no prior examples and is a widely used optimization idea~\cite{yuan2024evaluating, XX}. Although many studies~\cite{XX} have leveraged few-shot learning to learn from examples during inference to enhance LLM effectiveness, the type diversity and structure complexity nature of serverless application configurations makes it challenging to gather a set of examples that cover all configuration types. 

% Many studies~\cite{XX} have leveraged few-shot learning to enhance LLM effectiveness, allowing models to learn from examples during inference. However, the type diversity and structure complexity nature of serverless application configurations makes it challenging to gather a set of examples that cover all configuration types. Therefore, in this paper, we employ a zero-shot learning approach, which requires no prior examples and is a widely used optimization idea~\cite{yuan2024evaluating, XX}.

% In this section, we introduce our designed multi-dimensional constraints, which are based on the configuration characteristics of serverless applications. As discussed in Section~\ref{sec:background}, these configurations are summarized as the following key features:

%\wjf{configuration 组成成分去说，去掉关键词，core 组成}

Multi-dimensional constraints are designed based on the configuration characteristics of serverless applications. As introduced in Section~\ref{sec:example}, the constituent elements of a configuration file are diverse and encompass the following aspects:

\noindent $\bullet$ \underline{\textit{Resource Types}}: Serverless application configurations are primarily centered around defining resource types (e.g., lines 17 and 36 in Fig.~\ref{fig:configuration-example}). Resource types are core to establishing application execution settings. For instance, custom names such as ``BucketEventConsumer'' (line 16) are assigned to objects tied to specific resource types, such as ``AWS::Serverless::Function''. Moreover, resource type names are case-sensitive.

\noindent $\bullet$ \underline{\textit{Configuration Entries}}: Each resource type specifies diverse execution parameters, including language runtime and required resources for predefined events. These parameters are represented by configuration entries, such as \texttt{Runtime} on line 20 and \texttt{Events} on line 22 in Fig.~\ref{fig:configuration-example}.

% Resource types define executing setting including rich parameters, such as language runtime, memory size, and timeout, as well as the resources required for predefined events. These settings are represented by crucial configuration entries, e.g., \texttt{Handler} on line 19 and \texttt{Runtime} on line 20 in Figure~\ref{fig:configuration-example}.

\noindent $\bullet$ \underline{\textit{Values of Configuration Entries}}: Each configuration entry is assigned specific values, often governed by varied constraints. For example, the \texttt{Runtime} entry (line 20) has a set of allowed languages, e.g., ``python3.6'' and ``nodejs16.x'', while the \texttt{Bucket} entry (line 27) accepts only referenced objects.

% Each configuration entry is assigned varied values that adhere to specific constraints. For example, the \texttt{Runtime} entry on line 20 allows allocated different language runtime, such as ``python3.6''.  The \texttt{Bucket} entry on line 27 only allocates a referenced object.

\noindent $\bullet$ \underline{\textit{Entry Dependencies}}: Certain configuration entries depend on others. For example, \texttt{Name} from line 32 and \texttt{Value} from line 33 need to appear together in Fig.~\ref{fig:configuration-example}. These relationships are implicit and generally discovered by consulting documentation.

% a \textit{FunctionCode} entry~\cite{FunctionCode} within the ``AWS::Serverless::Function'' resource type relies on both the \texttt{Bucket} and \texttt{Key} entries to uniquely locate the code in S3, showing interdependencies. These relationships are implicit and generally discovered by consulting documentation.

% For instance, a developer may define a \textit{FunctionCode} entry~\cite{FunctionCode} in the resource type ``AWS::Serverless::Function''. This entry relies on the \texttt{Bucket} and \texttt{Key} configuration entries to uniquely identify the code's location in S3, underscoring the dependencies between these entries. Moreover, these kinds of dependencies are implicit and can only be discerned by delving into its documentation.

\noindent $\bullet$ \underline{\textit{Value Dependencies}}: Some values of configuration entries are interdependent across different resource types. For instance, the \texttt{RestApiId} entry for API event triggers depends on the object name value corresponding to the ``AWS::Serverless::Api'' resource. This shows how values can be linked across different resource types, showing extensive value dependencies. Such dependencies are common in configurations due to the collaboration between FaaS and BaaS.

% Some configuration entry values are interdependent. For example, the \texttt{RestApiId} entry for API event triggers depends on the object name value corresponding to the ``AWS::Serverless::Api'' resource. This demonstrates how values within certain entries can be linked across different resource types, highlighting dependencies among values. Moreover, value dependencies are common in configurations due to the collaboration of FaaS and BaaS.

% Based on these characteristics, we will define five dimensions of constraints to improve the LLM's effectiveness in identifying serverless application configurations: 

Based on these configuration characteristics, we design five dimensions of constraints (i.e., multi-dimensional constraints) to enhance the LLM's ability to identify serverless application misconfigurations:
\textbf{resource type constraint}, \textbf{entry constraint}, \textbf{value constraint}, \textbf{entry dependency constraint}, and \textbf{value dependency constraint}. Fig.~\ref{fig:goalprompt} shows their details. 

% We will explain them.

% Details are illustrated in Fig.~\ref{fig:goalprompt} and explained as follows. 

% Before explaining each constraint, we first introduce \wjf{Chain of Thought (CoT) technique~\cite{}}. \toolName employs the CoT technique for multi-dimensional constraints to improve the reasoning capability of the LLM. CoT is a reasoning approach in which the LLM generates intermediate steps or explanations that guide the problem-solving process, leading to more accurate and logical final answers. This technique helps the LLM break down complex tasks and improve its overall performance. We use ``category-by-category'' to conduct this reasoning approach.

Before explaining constraints, we introduce the Chain of Thought (CoT) technique~\cite{CoT, chu2023survey, li2023structured}. CoT is a reasoning strategy to guide the problem-solving process toward more accurate and logical conclusions. This technique breaks down complex tasks into smaller, manageable steps. A CoT-based prompt includes several intermediate natural language reasoning steps that describe how to solve the task step by step. Based on the principle of this technique, we design our CoT strategy for detecting misconfigurations of serverless applications by guiding LLMs to consider constraints in a ``category-by-category'' manner.

For \textbf{resource type constraint}, we describe it as follows: ``Check whether the resource type is currently supported by AWS SAM, search the following URL\footnote{Supported resource types: https://docs.aws.amazon.com/serverlessrepo/latest/devguide/list-supported-resources.html} to compare all supported AWS resources listed, noting the letter case.'' By providing a direct link to the official documentation, we enable \toolName to effectively identify and compare resource type names, with a particular focus on case sensitivity, a critical aspect in AWS SAM configurations.

% We provide a direct link to the official documentation to enhance the effective identification of resource types. This ensures that \toolName can compare names of resource types when considering case sensitivity, a critical factor in AWS SAM.

% Specifically, we describe \textbf{Resource Type Constraints} as: ``Check whether the resource type is currently supported by AWS SAM, search the following URL\footnote{Supported resource types: https://docs.aws.amazon.com/serverlessrepo/latest/devguide/list-supported-resources.html} to compare all supported AWS resources listed, noting the letter case.'' 
% To enhance the LLM's ability to accurately identify resource types, we provide a direct link to the official documentation, ensuring it can compare entries while considering case sensitivity, a critical factor in AWS SAM compliance.

% Specific description is shown in Listing

% Specifically, we describe \textbf{Resource Type Constraints} as ``Check whether the resource type is currently supported by AWS SAM, search the following URL\footnote{Supported resource types: https://docs.aws.amazon.com/serverlessrepo/latest/devguide/list-supported-resources.html} to compare all supported AWS resources listed, noting the letter case''. We provide a specified link from official documentation to help the LLM search supported resource types, ensuring check effectiveness. Case issues are considered because AWS SAM is sensitive to them.

For \textbf{entry constraint}, we design a three-step validation process to ensure the correctness of configuration entries. The first step checks the correctness of each entry in relation to its corresponding resource type. The second step checks the correctness of event-related entries. The third step ensures that all configuration entries follow the correct hierarchical structure. \toolName applies these checks using the CoT technique, following a ``step-by-step'' process. The three steps are as follows.  \textit{Step 1:} \toolName checks that each configuration entry exists under its respective resource type. This includes checking the entry's name for accuracy, paying particular attention to case sensitivity, and the use of singular or plural forms. \textit{Step 2:} For event-related entries, \toolName checks that configuration entries corresponding to each event source type are present. If any non-existent entries are given, \toolName flags them for review. \textit{Step 3:} \toolName checks the correct hierarchical structure of all configuration entries, with special attention to indentation. Misplaced or improperly indented entries may lead to errors, as they will not be recognized under the expected resource type.
This three-step validation process allows \toolName to systematically detect errors, ensuring comprehensive and accurate checks for configuration entries.

For \textbf{value constraint}, we describe it as follows: ``Check that the value type, constraints, and supported values of the configuration entry are correct, that the value representation is accurate, and that the value cannot be defined as null''. These constraints consider various aspects such as the correct data type, valid value ranges, and proper value formatting, ensuring that all values adhere to the required specifications.

% We design \textbf{Value Constraints} as follows: ``Check that the value type, constraints, and supported values of the configuration entry are correct, that the value representation is accurate, and that the value cannot be defined as null.''. These constraints encompass various aspects such as the correct data type, valid value ranges, and proper value formatting, ensuring that all values adhere to the required specifications.

% For \textbf{Value Constraints}, we describe it as: ``Ensure that the value type, constraints, and supported values for each configuration entry are correct, the value representation is accurate, and that the value cannot be null''. These constraints encompass various aspects such as the correct data type, valid value ranges, and proper value formatting, checking whether all values adhere to the required specifications.

% We define \textbf{Value Constraints} as follows: "Ensure that the value type, constraints, and supported values for each configuration entry are correct, the value representation is accurate, and that the value cannot be null." These constraints encompass various aspects such as the correct data type, valid value ranges, and proper value formatting, checking whether all values adhere to the required specifications.

For \textbf{entry dependency constraint}, we describe it as:  ``Check if there are dependencies between configuration entries, check that they are used in the correct way''. We also provide specific guidelines for validating dependencies, such as checking the accuracy of referenced resource types, ensuring required reference definitions are present, and confirming that required function entries are properly configured. 
% These checks help ensure that dependencies between entries are correctly established.

% Meanwhile, we provide some specific dependency descriptions to check the accuracy of referenced resource types, relevant required reference definitions, and unique function entries to guarantee proper configuration.

% We define \textbf{Entry Dependency Constraints} as: ``Check if there are dependencies between configuration entries, check that they are used in the correct way.'' Additionally, we provide specific dependency descriptions to validate the accuracy of referenced resource types, relevant required reference definitions, and unique function entries to guarantee proper configuration.

For \textbf{value dependency constraint}, we specify it as: ``Check if there is a dependency (possibly implicit) between the values of configuration entries, check that the usage is correct and that the relevant required reference definitions are given''. This constraint ensures that value dependency checks are comprehensive across the configuration, helping to maintain consistency and correctness in how values interact and depend on each other within the configurations.

% This ensures that checks of value dependencies are comprehensive across the configuration.

% Generally, LLMs are constrained by the number of input tokens they can process per query. However, in practice, the configuration files of serverless applications tend to not be relatively large. In our experimental evaluation, the maximum number of the configuration parameters of tested configuration files was more than 500. Based on this, we calculate the total number of tokens needed to describe multiple constraints. The resulting token count remains well within the input size limits of current LLMs.

% Generally, LLMs are limited by the number of input tokens they can process per query. However, in practice, the configuration files of serverless applications are not very large. In our experimental evaluation, the maximum number of configuration parameters in the tested files is 522. Based on this, we calculated the total number of tokens required to describe the multiple constraints. The resulting token count remains well within the input size limits of LLMs, ensuring that the models can process the entire configuration effectively.

\subsubsection{Customized Response}

% \begin{figure*}[t]
% \centering
%     \includegraphics[width=0.9\textwidth]{pic/ourresponse.pdf}
%     % \vspace{-1mm}
%     \caption{The generated content of \textit{Answer Requirement} module in \toolName.}
%     % \vspace{-1mm}
%     \label{fig:ouresponse}
%     % \vspace{-3mm}
% \end{figure*}

% Fig.\ref{fig:ouresponse} shows our customized response.

% the prompt part related to the response, which will reflect the specific answer requirements. It customizes both the content demand and format demand of the response to ensure effectiveness and relevance, shown in Fig.\ref{fig:ouresponse}.

We customize the LLMs' output by specifying both the content and format requirements for the responses, ensuring their effectiveness and relevance.
For the content demand, we aim to avoid receiving vague or uncertain answers that fail to explicitly identify configuration errors. To achieve it, we instruct the model with the directive: ``Please summarize the misconfigurations that are absolutely certain''. This ensures that only clear, deterministic errors are returned. Additionally, when applicable, we categorize the detected misconfigurations into specific groups, including ``Resource Type Errors,'' ``Configuration Entry Errors,'' ``Configuration Entry Value Errors,'' ``Entry Dependency Errors,'' and ``Value Dependency Errors''.

% Furthermore, we categorize the detected misconfigurations into specific groups when applicable, i.e., ``Resource Type Errors,'' ``Configuration Entry Errors,'' ``Configuration Entry Value Errors,'' ``Entry Dependency Errors,'' and ``Value Dependency Errors''.

% we aim to avoid receiving answers that do not explicitly identify configuration errors. To achieve this, we constraint the response with the instruction: ``Please summarize the misconfigurations that are absolutely certain''. This ensures that only deterministic errors are returned. Additionally, we categorize these detected misconfigurations by category, when present, as follows: ``Resource Type Errors,'' ``Configuration Entry Errors,'' ``Configuration Entry Value Errors,'' ``Entry Dependency Errors,'' and ``Value Dependency Errors.''

For the format demand, to eliminate redundant content that does not reveal specific misconfigurations from the raw output, we use delimiters: ``<START>'' and ``<END>'', to mark the required portion of the response. In \toolName, the desired output is enclosed within these markers, for example: ``<START> Resource Type Errors: ..., Value Dependency Errors: ... <END>''. This structured way ensures that only the relevant content is captured. During post-processing, \toolName employs regular expressions to extract the information between these markers efficiently. Although the model might generate additional text beyond the expected response, the use of locators allows for the seamless extraction of relevant content while discarding unnecessary text.

% to filter out any redundant content from the raw output, we use a locator pair, ``<START>'' and ``<END>'', to delimit the required portion of the response. In \toolName, the desired output is enclosed within these markers, such as: ``<START> Resource Type Errors: ..., Value Dependency Errors: ... <END>.'' This structure helps differentiate the relevant content from any extra text generated by the model. During post-processing, \toolName uses regular expressions to efficiently extract the information between the locators. While the model may still produce excess text beyond the expected response, the use of locators allows for easy extraction of the relevant content, with the rest discarded seamlessly.

\section{Experimental evaluation}\label{sec:evaluation}

To evaluate the effectiveness of \toolName in identifying misconfigurations within serverless applications, we present four research questions (Section~\ref{sec:rq}). To answer these questions, we detail the evaluation metrics (Section~\ref{sec:metric}), baselines for comparison (Section~\ref{sec:baseline}), evaluation dataset (Section~\ref{sec:datasets}), and experimental settings (Section~\ref{sec:environment}).

% Finally, we give evaluation results (Section~\ref{sec:results}).

% We propose the following research questions to assess the effectiveness of \toolName:
\subsection{Research Questions}\label{sec:rq}

\noindent $\bullet$ \textbf{RQ1:} How does the effectiveness of \toolName compared to traditional data-driven methods?

\noindent $\bullet$ \textbf{RQ2:} How effective is \toolName without considering our multi-dimensional constraints?

\noindent $\bullet$ \textbf{RQ3:} How does the non-determinism of LLMs influence the effectiveness of \toolName?

% \wjf{RQ3: How does the non-determinism of LLMs influence the effectiveness of \toolName?}

\noindent $\bullet$ \textbf{RQ4:} How does the generalization capability of \toolName when using different LLMs?

\subsection{Evaluation Metrics}~\label{sec:metric}
We use $precision$, $recall$, and $F1$-$score$ as evaluation metrics to compare \toolName against the baseline methods at the configuration parameter level, i.e., configuration entries or values. We check whether the detection approach can accurately determine the validity of each configuration parameter within the configuration file.
% These metrics provide a quantitative basis for comparing approach's effectiveness in identifying misconfigurations, as explained as follows:
% These metrics are explained in detail as follows. 
$precision$ measures the proportion of correctly identified misconfigured parameters among all parameters flagged as misconfigured. 
% In other words, it evaluates how many of the flagged parameters are truly misconfigured. 
$recall$ quantifies the ability of the approach to detect actual misconfigurations by calculating the proportion of true misconfigured parameters that are correctly identified. $F1$-$score$ provides a balanced measure that accounts for the significance of both false positives and false negatives. 
These metrics are calculated through True Positives (TP), False Positives (FP), True Negatives (TN), and False Negatives (FN), explained in Table~\ref{tab:metrics}. $precision$ = $\frac{TP}{TP + FP}$, $recall$ = $\frac{TP}{TP + FN}$, and $F1$-$score$ = $2 \times \frac{precision \times recall}{precision + recall}$. Values range from 0\% to 100\%, with scores closer to 100\% indicating greater effectiveness.

\begin{table*}[t]
\footnotesize
 \caption{The explanation of TP, FP, TN, and FN in our scenario.}
 \vspace{-3mm}
    \label{tab:metrics}
    \begin{tabular}{l|l}
    \hline
   \textbf{TP} & A misconfigured parameter correctly identified as misconfigured \\ \hline
   \textbf{FP} & A correctly configured parameter mistakenly flagged as misconfigured \\ \hline
   \textbf{TN} & A correctly configured parameter accurately recognized as valid \\ \hline
   \textbf{FN} & A misconfigured parameter that is overlooked or incorrectly classified as valid \\ \hline
\end{tabular}
\vspace{-3mm}
\end{table*}

\subsection{Baseline Methods}~\label{sec:baseline}
% We implement two types of baselines to compare the effectiveness of misconfiguration detection approaches of serverless applications. First, currently, no methods are tailored specifically for detecting misconfigurations in serverless applications. We will adopt the principles of existing data-driven techniques that were widely designed by previous configuration work to design a misconfiguration detection approach in the serverless application scenario. In addition, we also implement a simple LLM-based approach. These baselines are as follows:
% Since there are currently no methods specifically tailored for detecting misconfigurations in serverless environments, we first adopt principles from existing data-driven techniques commonly used in previous configuration work~\cite{zhang2014encore, santolucito2016probabilistic, zhou2023drive, santolucito2017synthesizing}. 
% We implement two types of baselines to evaluate the effectiveness. 
% Given the lack of approaches specifically tailored for detecting misconfigurations in serverless computing, we first draw on principles from established data-driven techniques used in prior configuration studies~\cite{zhang2014encore, santolucito2016probabilistic, zhou2023drive, santolucito2017synthesizing}. By adapting these methods, we create a data-driven baseline tailored to the characteristics of serverless applications. Additionally, we introduce a straightforward LLM-based baseline as a second comparison, which is not consider our designed constraints.
We implement two types of baselines to evaluate effectiveness. Given the lack of approaches specifically tailored for detecting misconfigurations in serverless computing, we first draw on principles from established data-driven techniques used in prior configuration studies~\cite{zhang2014encore, santolucito2016probabilistic, zhou2023drive, santolucito2017synthesizing}. By adapting these methods, we create a data-driven baseline suited to the characteristics of serverless applications. Additionally, we introduce a straightforward LLM-based baseline as a second comparison, which does not consider our designed constraints.

% By leveraging their implementation idea, we design a data-driven baseline approach that can be applied to the unique context of serverless applications. In addition, we also present a simple LLM-based approach as a second baseline to compare our designed \toolName.

\noindent $\bullet$ \textbf{Baseline 1:} \underline{Data-driven method (DD method}). We implement a data-driven approach for serverless applications by learning configuration patterns from a dataset of configuration files. As no existing dataset specifically focuses on serverless application configurations, we collect our data from the AWS Serverless Application Repository (SAR)~\cite{SAR}, an official repository for serverless applications where each application is packaged with an AWS SAM template and links to relevant configuration files. We include all configuration files associated with serverless applications that have been successfully deployed at least once as of August 18, 2023, which is the date we collected this dataset. This results in a collection of 701 configuration files across 658 serverless applications, with some links providing multiple configuration files representing distinct configurations.
Given the correctness of ensuring the dataset, we conduct a careful manual review of the configuration files. This review was performed by the first two authors, who have a background in cloud computing. Identified issues were discussed and resolved with consensus among the authors. To assess the consistency of independent labeling, we employ Cohen's Kappa ($\kappa$)~\cite{cohen1960coefficient}, a widely used metric for measuring inter-rater agreement. The resulting $\kappa$ value of 0.916 indicates an almost perfect agreement and a reliable labeling procedure~\cite{landis1977measurement}.

% However, it is hard to completely ensure the correctness of the collected dataset. To end, we manually and carefully check these configuration files through the first two authors with a background in cloud computing knowledge. The determined problems are modified, which are agreed upon among the authors. To measure the inter-rater agreement level of the authors during the independent labeling, we use Cohen's Kappa ($\kappa$)~\cite{cohen1960coefficient}, which is the most widely-used agreement evaluation metric. The value of $\kappa$ is 0.916, indicating an almost perfect agreement and a reliable labeling procedure~\cite{landis1977measurement}.

% Using this dataset, we mine configuration patterns, including common resource types, configuration entries, values, and dependencies among these entries and values. To facilitate the mining process, we first standardize the configuration files into uniform representations. Object names for various resource types are identified, and we replace customized object names with standardized labels (e.g., the placeholder ``PH+resource type'') for configuration entries and values. From there, we mine the configuration patterns for used resource types, entries, and values.
Using this dataset, we learn configuration patterns, focusing on common resource types, configuration entries, values, and dependencies among entries and values. To streamline this process, we first standardize the configuration files into a uniform representation. Object names for various resource types are identified, with object names replaced by standardized labels (e.g., a placeholder like ``PH+resource type'') for consistency across configuration entries and values. Leveraging this standardized dataset, we extract the used resource types, entries, and values.

To detect dependencies among both entries and values, we apply association rule mining techniques~\cite{xu2009detecting, yuan2011context}. Specifically, we use the FP-Growth algorithm~\cite{han2000mining}, which is known for its scalability. We need to set a support threshold for frequent itemsets using
the formula $\alpha \times len$, where $len$ represents the total number of configuration files, a deterministic value, and $\alpha$ is a percentage that indicates the desired mining granularity. Leveraging mined frequent itemsets, we generate association rules by utilizing traversal way and dividing items into left and right sets, where items in the right set must appear if those in the left set are present. These rules reveal the configuration dependencies. If the tested file contains all items in a left set, this approach checks whether it includes the corresponding items in the right set. If any items are missing, it reports them.

\begin{figure}[t]
\centering
    \includegraphics[width=0.95\textwidth]{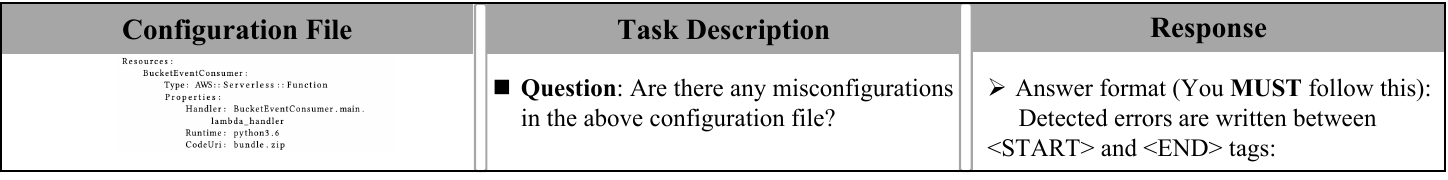}
    \vspace{-3mm}
    \caption{The prompt of BL method.}
    % \vspace{-1mm}
    \label{fig:blmprompt}
    \vspace{-3mm}
\end{figure}

\noindent $\bullet$ \textbf{Baseline 2}: \underline{Basic LLM method (BL method)}. It is designed using a straightforward prompt that does not take our multi-dimensional constraints into account. This prompt contains the configuration file content followed by a task description. Similarly, the output is enclosed within a locator pair, ``<START>'' and ``<END>'', to delimit the required response. This prompt is shown in Fig.~\ref{fig:blmprompt}.

% The experiment utilizes a basic prompt structure consisting of the configuration file content followed by the question. Similarly, the output results are enclosed within the locator pair, ``<START>'' and ``<END>'', to delimit the required portion of the response.
% The specific prompt is shown in Figure~\ref{fig:blmprompt}.

% ``Are there any misconfigurations in the above configuration file?''

\subsection{Evaluation Dataset}\label{sec:datasets}

We conduct experimental evaluations on a dataset comprising three types of configurations. The first type includes error-free configurations, enabling us to evaluate true negatives and false positives in detection. The second type contains configurations with real-world errors, allowing for the assessment of true positives and false negatives. Although this second type is somewhat free of data leakage concerns of LLMs, we include a third type to strengthen the validity of our conclusions. The third type consists of configurations with injected errors, which are not exposed to LLMs during training, thereby eliminating data leakage concerns. By utilizing these diverse configurations, we can achieve a valid evaluation.

\noindent $\bullet$ \underline{\textit{Configurations without Errors (26).}} We manually collect configuration files that have been successfully executed without errors. This data is separate from the one used to mine configuration patterns in the data-driven approach. 

We collect real-world configuration cases from GitHub. GitHub issues provide rich information, including developer discussions and related code or configuration fragments. 
% To collect configuration cases specific to AWS serverless computing, 
We conduct the following steps. First, on July 2, 2024, the date we collected this data, we searched GitHub using the keywords ``AWS,'' ``serverless,'' and ``configuration,'' which yielded more than 8,000 relevant configuration-related issues. We then manually reviewed these issues to extract correct configuration fragments from the problematic cases—a time-consuming and challenging process. To facilitate this task, the first two authors jointly review the configurations. Initially, they filter through the configuration fragments by searching for terms including ``successful,'' ``successfully,'' and ``it works'' within the issues to identify correct configurations. For the fragments that matched, they conducted a manual verification process to ensure that the configurations were indeed error-free.
Over two months, the two authors identified 52 configuration fragments that met our criteria. These error-free real-world configuration fragments are divided into two sets: 26 (naming from case 1 to case 26) are used to evaluate error-free configurations, while the remaining 26  (naming from case 27 to case 52) are reserved for generating configurations with injected errors, which is explained in detail later.

\noindent $\bullet$ \underline{\textit{Real-world Misconfigurations (58).}} 
To evaluate the effectiveness of approaches in identifying real-world misconfigurations in serverless applications, we construct a relevant dataset by mining real-world configuration issues from GitHub. These issues need to contain clearly identified root causes as ground truths, enabling us to accurately assess the effectiveness of detection results.

% we construct a relevant dataset. We mine real-world configuration problems from GitHub. These problems include the determined root causes, helping to determine the correctness of detection results of \toolName.

The selection process is as follows: First, we use the same keywords (i.e., ``AWS,'' ``serverless,'' and ``configuration'') to search for relevant issues on GitHub on July 2, 2024. Next, we identify satisfied issues based on the following criteria: (i) the issue is marked as closed, indicating that it has been resolved; (ii) the issue includes a configuration fragment based on AWS SAM for analysis; and (iii) the discussion concludes with a clearly identified root cause of the problem.
Using these criteria, we select 58 real-world configuration problems encountered in serverless applications, surpassing the scale of prior studies on configuration-related research~\cite{zhang2014encore, yuan2011context}.

To ensure the accuracy of the configuration errors to be detected, we meticulously review each real-world configuration file in conjunction with its identified root cause. During this process, we also manually identify and address any potential configuration issues (e.g., outdated runtime) that could influence the evaluation.

% Out of these 70 misconfigurations, 6 involve resource type violations, 31 pertain to configuration entry violations, 19 concern suspicious values, 2 relate to configuration dependency violations, and the remaining 13 are associated with special constraint violations.

% \wjf{to ensure no other errors in each configuration file}

% (3) \underline{\textit{Injected Misconfigurations.}}. We construct a dataset of injected misconfigurations. We aim to generate various erroneous configurations on the correct configuration files.
% We use selected 26 real-world and correct configuration files collected from the first dataset in this section. Then, we generate misconfigurations of different types for these selected configuration files. There are generation rules from previous studies on misconfigurations~\cite{li2018confvd, li2021challenges, xu2013not, sun2020testing}, which violate the constraints of configurations. 
% We use existing rules and extend specific misconfiguration generation rules for serverless application configurations, as shown in Table~\ref{tab:misconfigurationinject}. Through matching subcategories, we randomly generate misconfigurations for each selected correct configuration file to form a new configuration file to be detected. Finally, we generate 26 configuration files with injected misconfigurations.

\begin{table*}[t]
\footnotesize
 \caption{Misconfiguration generation rules (we use generation rules from previous work~\cite{li2018confvd, li2021challenges, xu2013not, sun2020testing} and customize them in our scenario.)}
 \vspace{-3mm}
    \label{tab:misconfigurationinject}
    \begin{tabular}{c|c|p{4.8cm}|p{4.5cm}}
    \hline
    \textbf{Category} & \textbf{Subcategory} & \textbf{Specification} & \textbf{Generation Rules}\\
    \hline
        
       \multirow{2}{*}{Syntax} & \tabincell{c} {Resource \\type} & Value set = \{AWS::Serverless::Function, AWS::Serverless::Api, ...\}  & Generate a resource type that does not belong to the value set  \\ \cline{2-4}
        
         & \tabincell{c} {Entry} & Value set = \{entry1, entry2, ...\}, specific entries are used in a certain resource type & Generate an invalid entry for a resource type \\ \cline{1-4}
    
    \multirow{3}{*}{Range} & \tabincell{c} {Basic \\numeric} & Valid range constrained by data type & Generate values outside the valid range (e.g., max value+1)  \\ \cline{2-4}
    % & Bool & Options, value set = \{``TRUE'', ``FALSE''\}  & Generate a value that does not belong to the value set \\ \cline{2-4}
    & \tabincell{c} {Enum}  & Options, value set = \{enum1, enum2, ...\}, specific values are used in a certain configuration entry & Generate a value that does not belong to set \\ \cline{1-4}

    \multirow{2}{*}{Dependency}& \tabincell{c} {Entry \\relationship} & $(P_1, V, \diamond) \mapsto P_2, \diamond \in \{>, \geq, =, \neq, <, \leq, occurrence\}$  & Generate invalid entry relationships for configuration entries $(P_1, V, \neg \diamond)$ \\ \cline{2-4}
    & \tabincell{c} {Value \\relationship}   & $(P_1, P_2, \diamond), \diamond \in \{>, \geq, =, \neq, <, \leq, occurrence\}$ & Generate invalid value relationship for configuration entry values $(P_1, P_2, \neg \diamond)$  \\ \cline{2-4}
    \hline
\end{tabular}
\vspace{-3mm}
\end{table*}

\noindent $\bullet$ \underline{\textit{Injected Misconfigurations (26).}}
We construct injected misconfigurations by generating various errors in the correct configuration files. To achieve this, we use 26 error-free configuration files named from case 27 to case 52. Misconfigurations of different types are then generated, following misconfiguration generation rules from prior studies~\cite{li2018confvd, li2021challenges, xu2013not, sun2020testing, lian2023configuration}. Prior studies~\cite{li2018confvd, lian2023configuration} showed that these rules can cover most configurations.
In addition to utilizing existing rules, we extend specific misconfiguration generation rules tailored to serverless application configurations, as outlined in Table~\ref{tab:misconfigurationinject}. For each selected configuration file, we randomly sample a configuration parameter that aligns with the subcategories in Table~\ref{tab:misconfigurationinject} and generate invalid configurations, creating a new erroneous configuration file for detection.
% For each selected configuration file, we randomly sample  configuration parameter that can match subcategories in Table~\ref{tab:misconfigurationinject}, and generate invalid configurations to create a new, erroneous configuration file for detection. 
In total, we generate 26 configuration files with injected misconfigurations for evaluation.

\begin{figure}[t]
\centering
    \includegraphics[width=0.93\textwidth]{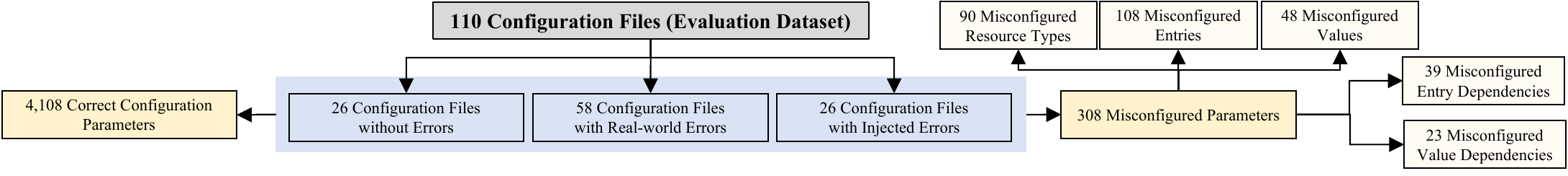}
    \vspace{-3mm}
    \caption{The Details of Evaluation Dataset.}
    % \vspace{-1mm}
    \label{fig:evaluationdataset}
    \vspace{-3mm}
\end{figure}

Our evaluation dataset contains 110 configuration files with corresponding ground-truth answers. Fig.~\ref{fig:evaluationdataset} shows its details. Of these, 26 are error-free configuration files, 58 contain real-world errors, and 26 have injected errors. Across all configuration parameters, there are 4,108 correct configuration parameters and 308 misconfigured ones. Among the misconfigured parameters, 90 involve incorrect resource types, 108 have misconfigured entries, 48 contain incorrect values, 39 exhibit entry dependency issues, and 23 have value dependency issues.
% Our evaluation dataset comprises 110 configuration files with corresponding ground-truth answers. Its details are shown in Fig.~\ref{fig:evaluationdataset}. 26 are configuration files without errors, 58 are configuration files with real-world errors, and 26 are configuration files with injected errors. In all configuration parameters, there are 4108 correct configuration parameters and 308 misconfigured parameters. For misconfigured parameters, 101 are involved in misconfigured resource types, 98 are misconfigured entries, 44 are misconfigured values, and 65 are misconfigured dependencies.
We analyze the detection results across all configuration parameters to obtain TP, FP, TN, and FN. We then calculate $precision$, $recall$, and $F1$-$score$ to evaluate the effectiveness of the detection.

% To evaluate the effectiveness of \toolName compared to baseline methods, we calculate the $precision$, $recall$, and $F1$-$score$ for their detection results.

% In total, our evaluation dataset contains 110 configuration files with corresponding ground-truth answers. By calculating $precision$, $recall$, and $F1$-$score$ of their detection results, we compare the approach effectiveness of \toolName and baselines.

\subsection{Experimental Settings}\label{sec:environment}

We introduce our parameter settings, experimental repetitions, and experimental environment.

\noindent \textbf{Parameter Settings.} For \textbf{RQ1}, the compared DD method needs to specify a frequent threshold, $\alpha$. We experiment with various threshold levels: low (1\%), medium (3\% and 5\%), and high (10\%). A lower threshold corresponds to a lower support value, enabling the discovery of more dependencies. For comparisons with \toolName, we use a default $\alpha$ value of 5\%. Experimental results also show that 5\% is optimal for achieving the best effectiveness results in DD methods.
We also report results for both \toolName and DD method across other thresholds. For \textbf{RQ2}, we compare \toolName with the BL method, both of which leverage LLMs. We select ChatGPT-4o as the default LLM due to the widespread use and outstanding performance of ChatGPT in recent research~\cite{yuan2024evaluating, lian2023configuration}.
A crucial parameter of LLMs is the temperature, which controls the level of randomness in the generated responses. To ensure reproducibility and consistency, we follow the previous work~\cite{yin2024multitask, xu2024divlog, hadadi2024anomaly, chen2024automatic-paper7} to set the temperature to 0 for all identical queries.
For \textbf{RQ3}, there are no specific parameters to be set.
For \textbf{RQ4}, we evaluate the generalization capability of \toolName across various LLMs, excluding ChatGPT-4o. Specifically, we utilize an open-source model, Llama 3.1 (405B) Instruct Turbo, and a proprietary model, Gemini 1.5 Pro. These models are among the top-ranked LLMs~\cite{LLMrank}.
% Llama 3.1 (405B) Instruct Turbo, an open-source model, and Gemini 1.5 Pro, a proprietary model. 
% Both are popular and widely used in current LLM research. 
As with RQ2, we set the temperature of LLMs to 0 to maintain consistent outputs across repeated queries.

\noindent \textbf{Experimental Repetitions.} For experiments involving stochastic processes, we follow established best practices~\cite{xu2024divlog, hadadi2024anomaly}, repeating each experiment five times and reporting the mean evaluation metrics to reduce the impact of random variations.

% \textbf{Experimental Environment.}  We conducted our experiments on an Ubuntu 18.04.4 LTS server equipped with an Intel Xeon (R) 4-core processor and 24GiB of memory. The LLMs were accessed via their respective APIs. Although all methods are implemented in Python, their ability to detect configuration issues operates independently of the programming language used, making them adaptable as end-to-end solutions. For experiments involving stochastic processes, we performed five repetitions and reported the median results of evaluation metrics, following established best practices~\cite{jiang2024large, jiang2024lilac}, to mitigate the impact of random variation.

\noindent \textbf{Experimental Environment.} Our experiments were conducted on an Ubuntu 18.04.4 LTS server with an Intel Xeon (R) 4-core processor and 24 GiB of memory. The LLMs were accessed through their respective APIs. While all methods are implemented in Python, their misconfiguration detection capabilities are independent of the underlying programming language. 

% For experiments involving stochastic processes, we followed established best practices~\cite{jiang2024large, jiang2024lilac}, repeating each experiment five times and reporting the median evaluation metrics to reduce the impact of random variations.
\section{Evaluation Results}~\label{sec:results}

This section gives and discusses the results of each research question.

\subsection{RQ1: Effectiveness of \toolName and Data-Driven Method (DD method)}

% This section explores the effectiveness of \toolName compared to the DDM method. Results show that \toolName is highly effective. Table~\ref{tab:RQ1} shows the evaluation metric results of \toolName and DDM in detecting misconfigurations of serverless applications. Specifically, \toolName provides the precision with XX, recall with XX, and F1 with XX. DDM with default threshold
% only obtains XX of precision, XX of recall, and XX of F1. Compared to DDM, \toolName improves the precision by XX percentage points, the recall by XX percentage points, and the F1 by XX percentage points. This indicates the effectiveness of \toolName. We further analyze the reasons that DDM produces low-effectiveness results. DDM has a low precision, i.e., XX, which is due to a high FP value. This means many correctly configured parameters are mistakenly flagged as misconfigured. DDM is a data-driven method that learns configuration patterns from historical data, including previously supported configurations, and may not include newly supported values or even configurations that do not appear. Thus, this kind of method makes it difficult to identify some correct configurations that actually supported.

\begin{table*}[t]
\footnotesize
 \caption{RQ1: Results about \toolName and DD method.}
 \vspace{-3mm}
    \label{tab:RQ1}
    \begin{tabular}{l|c|c|c}
    \hline
    \textbf{Methods} &  $precision$ & $recall$ & $F1$-$score$\\
    \hline
    DD method with 5\% threshold (default) & 19.06\% & 70.78\% & 30.03\% \\ \hline
    \textbf{\toolName} (vs DD method with 5\% threshold) & 72.88\% ($\uparrow$ 53.82\%) & 88.18\% ($\uparrow$ 17.40\%) & 79.75\% ($\uparrow$ 49.72\%) \\ \hline
    \hline
    DD method with 10\% threshold & 17.70\% & 64.61\% & 27.79\% \\ \hline
    DD method with 3\% threshold & 18.83\% & 70.13\% & 29.69\% \\ \hline
    DD method with 1\% threshold & 18.85\% & 70.45\% & 29.75\% \\ \hline
    
    % \hline
\end{tabular}
\vspace{-3mm}
\end{table*}

This section explores the effectiveness of \toolName in comparison to the DD method. \toolName has a significant advantage in the effectiveness aspect. 
Table~\ref{tab:RQ1} presents their results in detecting misconfigurations in serverless applications. Specifically, \toolName achieves a $precision$ of 72.88\%, $recall$ of 88.18\%, and $F1$-$score$ of 79.75\%. In contrast, the DD method, with its default threshold of 5\%, only reaches a $precision$ of 19.06\%, $recall$ of 70.78\%, and $F1$-$score$ of 30.03\%. \toolName outperforms the DD method, increasing $precision$ by 53.82 percentage points, $recall$ by 17.40 percentage points, and $F1$-$score$ by 49.72 percentage points, showing its superior effectiveness.

We investigate why the DD method produces less effective results. One major issue is its low $precision$ (19.06\%) and $F1$-$score$ (30.03\%). We further observe TP, FN, FP, and TN values obtained by the DD method across all configuration parameters, as shown in Table~\ref{tab:rq1tpfnfptn}. 
Results show that the FP value is 926, indicating that 22.54\% of the 4,108 correct configuration parameters are mistakenly flagged as misconfigurations. In contrast, on average, \toolName misclassifies only 2.48\% of correct configuration parameters as misconfigurations. Thus, the low effectiveness of the DD method is attributed to high false positives.
As a data-driven approach, the DD method learns configuration patterns based on historical data, which mainly includes previously used configurations. This reliance makes it difficult to accurately identify configurations that are either rare or newly supported, resulting in numerous false positives. Thus, the DD method fails to detect some valid configurations that are indeed supported, leading to its low $precision$ and $F1$-$score$.

\begin{table*}[t]
\begin{threeparttable}
\footnotesize
 \caption{RQ1: Results\tnote{*} of TP, FN, FP, and TN for DD method and \toolName.}
 \vspace{-0.5mm}
    \label{tab:rq1tpfnfptn}
    \begin{tabular}{|l|c|c|c|c|}
    \hline
      \multirow{2}{*}{\textbf{Methods}} & \multicolumn{2}{c|}{308 misconfigured parameters} &  \multicolumn{2}{c|}{4,108 correct configuration parameters} \\ \cline{2-5}
         &  \textbf{TP}  & \textbf{FN} & \textbf{FP} & \textbf{TN}  \\
    \hline
    DD method (default) & 218 (70.78\%) & 90 (29.22\%) & \textbf{926 (22.54\%)} & 3,182 (77.46\%)  \\ \hline
    \toolName (default) & 272 (88.31\%) \checkmark &	36 (11.69\%) \checkmark & 102 (2.48\%) \checkmark & 4,006 (97.52\%) \checkmark  \\ \hline

\end{tabular}
  \begin{tablenotes}
     \item[*] Higher TP and TN are preferable, while lower FN and FP are desired.
  \end{tablenotes}
   \end{threeparttable}
   \vspace{-3mm}
\end{table*}

We also compare the effectiveness of the DD method under different thresholds $\alpha$: 10\%, 3\%, and 1\%, with the results presented in Table~\ref{tab:RQ1}. As $\alpha$ decreases from 10\% to 1\%, the evaluation metrics show improvement. Specifically, $precision$ increases from 17.70\% to 18.85\%, $recall$ rises from 64.61\% to 70.45\%, and $F1$-$score$ improves from 27.79\% to 29.75\%. To further explore the reasons for their changes, we give TP, FN, FP, and TN results of the DD method under different thresholds, as shown in Table~\ref{tab:rq1tpfnfptnforDD}. 
The primary reason for improvements is that lower $\alpha$ mines more dependencies among entries or values. This enables the accurate identification of a larger number of misconfigured parameters. Specifically, the TP value for the DD method at a 10\% threshold is 199, whereas at a 1\% threshold, it increases to 217. This improvement leads to a higher $recall$, increasing from 64.61\% to 70.45\%.
% The primary reason for improvements is that lower $\alpha$ thresholds will uncover more dependencies among configuration entries or values. This allows for more misconfigured parameters to be accurately identified. TP of the DD method with 10\% is 199, while the TP of the DD method with 1\% is 217. This leads to higher $recall$, i.e., from 64.82\% to 70.68\%. 
% However, a lower threshold also increases the risk of generating more redundancy and invalid dependency relationships, resulting in some correctly configured parameters being mistakenly flagged as misconfigured. This can be obtained from FP values. FP of the DD method with 10\% is 925, while the FP of the DD method with 1\% is 934. Consequently, $precision$ does not show as much improvement, i.e., 17.70\% to 18.85\%. 
However, a lower $\alpha$ also increases the risk of generating potentially invalid dependencies, resulting in correctly configured parameters being mistakenly flagged as misconfigurations. This is evident from the FP values: the FP value for the DD method at a 10\% threshold is 925, while at a 1\% threshold, it increases to 934. As a result, $precision$ shows only a modest improvement, from 17.70\% to 18.85\%.
For $F1$-$score$, lowering $\alpha$ enhances the effectiveness of the DD method, reaching a value of 29.75\%. However, it still significantly lags behind the 79.75\% achieved by \toolName.
% Nevertheless, this value still falls short of $F1$-$score$ (79.67\%) achieved by \toolName.

% In addition, we compare the effectiveness of DDM under different frequent thresholds: 10\%, 3\%, and 1\%. Their results are shown in Table~\ref{tab:RQ1}. Evaluation metric values have improvements as the frequent thresholds decrease from 10\% to 1\%. Specifically, precision values improve from 17.70\% to 18.85\%, recall values are from 64.82\% to 70.68\%, and F1-score values are from 27.81\% to 29.77\%. The main reason for these improvements is that using a low threshold will mine more dependencies among configuration entries or values. This means more misconfigured parameters are correctly identified as misconfigured, producing a higher recall. However, using a low threshold increases the risk of producing more potential redundancy and invalid relationships. This makes correctly configured parameters be mistakenly flagged as misconfigured, thus precision values do not improve much. From the view of F1-score values, using a lower threshold can improve the effectiveness of DDM, producing 29.77\% of the value. However, this value still does not exceed the F1-score value (XX) of \toolName.

In addition, we observe that a threshold of 5\% for the DD method yields superior results compared to 1\%, 3\%, and 10\%, suggesting that 5\% is an optimal threshold for the data-driven method in this scenario. In the threshold of 5\%, the FP-growth algorithm can effectively mine relationships without losing valid dependencies or generating an excessive number of invalid dependencies. However, even at 5\%, the effectiveness of the DD method remains significantly lower than that of \toolName, with particularly low $precision$ and $F1$-$score$. 

% This ineffectiveness can be attributed to the inherent limitations of the data-driven approach, which relies on learning existing configuration patterns from historical data, leading to less effective outcomes.

% We also find that the effectiveness of DDM at 5\% threshold is superior to the effectiveness of DDM at other thresholds (1\%, 3\%, and 10\%). This indicates that 5\% is an appropriate threshold to select in this scenario. FP-growth algorithms mine relationships without losing valid dependencies or generating too many redundant invalid dependencies. However, even at this optimal threshold, the DDM method still produces far less effectiveness than \toolName, leaving low precision and F1-score values. Thus, the low effectiveness of the DDM method sources from inherent limitation of data-driven way, which requires learning existing knowledge from historical data.

\begin{table*}[t]
\footnotesize
 \caption{RQ1: Results of TP, FN, FP, and TN for DD method with different thresholds $\alpha$.}
 \vspace{-3mm}
    \label{tab:rq1tpfnfptnforDD}
    \begin{tabular}{|l|c|c|c|c|}
    \hline
      \multirow{2}{*}{\textbf{Methods}} & \multicolumn{2}{c|}{308 misconfigured parameters} &  \multicolumn{2}{c|}{4,108 correct configuration parameters} \\ \cline{2-5}
         &  \textbf{TP}  & \textbf{FN} & \textbf{FP} & \textbf{TN}  \\
    \hline
    DD method with 10\% threshold & 199 & 109 & 925 & 3,183 \\ \hline
    DD method with 3\% threshold & 216  & 92 & 931 & 3,177  \\ \hline
    DD method with 1\% threshold & 217 & 91 & 934 & 3,174  \\ \hline
\end{tabular}
\vspace{-3mm}
\end{table*}

\finding{
% \toolName achieves a $precision$ of 71.83\%, $recall$ of 87.01\%, and $F1-score$ of 79.94\%. 
% \toolName outperforms traditional data-driven methods that have the best results across all evaluation metrics, achieving significant improvements of 52.77 percentage points in $precision$, 16.00 percentage points in $recall$, and 49.89 percentage points in $F1-score$. These results suggest the superior effectiveness of our approach compared to traditional approaches.
\toolName achieves a $precision$ of 72.88\%, $recall$ of 88.18\%, and $F1$-$score$ of 79.75\%, surpassing data-driven methods across all metrics. It shows significant improvements, with increases of 53.82 percentage points in $precision$, 17.40 percentage points in $recall$, and 49.72 percentage points in $F1$-$score$. These results suggest the high effectiveness of \toolName.
}

\subsection{RQ2: Effectiveness of \toolName and Basic LLM-based Method (BL method)}

\begin{table*}[t]
\footnotesize
 \caption{RQ2: Results about \toolName and BL method using the default LLM (ChatGPT-4o).}
 \vspace{-3mm}
    \label{tab:RQ2result}
    \begin{tabular}{l|c|c|c||p{1.8cm}|p{1.3cm}|p{1.3cm}|p{1.3cm}}
    \hline
    \textbf{Baseline} &  $precision$ & $recall$ & $F1$-$score$ & \textbf{Our Approach} &  $precision$ & $recall$ & $F1$-$score$\\
    \hline
    
    BL method & 51.65\% & 65.00\%  & 57.55\% & \tabincell{c} {\textbf{\toolName} \\ (vs BL method)} &  \tabincell{c} {72.88\% \\ ($\uparrow$ 21.23\%)} &  \tabincell{c}{88.18\% \\ ($\uparrow$ 23.18\%)} &  \tabincell{c}{79.75\%\\ ($\uparrow$ 22.20\%) }\\ \hline

    % \hline
\end{tabular}
\vspace{-3mm}
\end{table*}

\begin{table*}[t]
\begin{threeparttable}
\footnotesize
 \caption{RQ2: Results of TP, FN, FP, and TN for BL method and \toolName, on average.}
 % \vspace{-3mm}
    \label{tab:rq2tpfnfptn}
    % \vspace{-3mm}
    \begin{tabular}{|l|c|c|c|c|}
    \hline
      \multirow{2}{*}{\textbf{Methods}} & \multicolumn{2}{c|}{308 misconfigured parameters} &  \multicolumn{2}{c|}{4,108 correct configuration parameters} \\ \cline{2-5}
         &  \textbf{TP}  & \textbf{FN} & \textbf{FP} & \textbf{TN}  \\
    \hline
    BL method (default) & \textbf{200 (64.94\%)} & 107 (34.74\%) & 188 (4.58\%) & 3,920 (95.42\%)  \\ \hline
    \toolName (default) & 272 (88.31\%) \checkmark &	36 (11.69\%) \checkmark & 102 (2.48\%) \checkmark & 4,006 (97.52\%) \checkmark  \\ \hline

\end{tabular}
  \begin{tablenotes}
     \item[*] Higher TP and TN are preferable, while lower FN and FP are desired.
  \end{tablenotes}
   \end{threeparttable}
   \vspace{-3mm}
\end{table*}

% \begin{table*}[t]
% \footnotesize
%  \caption{RQ2 and RQ3: Results about \toolName and BL method using different LLMs, where Llama is open source and Gemini is proprietary.}
%     \label{tab:RQ23}
%     \begin{tabular}{l|c|c|c||p{2cm}|p{1.4cm}|p{1.4cm}|p{1.4cm}}
%     \hline
%     \textbf{BL Method} &  $precision$ & $recall$ & $F1$-$score$ & \textbf{Our Approach} &  $precision$ & $recall$ & $F1$-$score$\\
%     \hline
    
%     BL (GPT-4o) & 51.44\% & 65.26\%  & 57.35\% & \tabincell{c} {\textbf{\toolName} \\(GPT-4o) (vs BL)} &  \tabincell{c} {71.83\% \\ ($\uparrow$ 20.39\%)} &  \tabincell{c}{87.01\% \\ ($\uparrow$ 21.75\%)} &  \tabincell{c}{79.94\%\\ ($\uparrow$ 22.59\%) }\\ \hline
%     \hline
%     \hline
%     BL (Llama) & 47.47\% & 57.79\%  & 53.54\% & \tabincell{c} {\textbf{\toolName}\\ (Llama) (vs BL)} & \tabincell{c} {70.29\% \\ ($\uparrow$ 22.83\%)} &  \tabincell{c}{79.80\% \\ ($\uparrow$ 22.01\%)} &  \tabincell{c}{74.89\%\\ ($\uparrow$ 21.35\%) }\\ \hline
%     \hline
%     BL (Gemini) & 44.21\% & 21.90\%  & 29.33\% & \tabincell{c} {\textbf{\toolName} \\(Gemini) (vs BL)} & \tabincell{c} {68.59\% \\ ($\uparrow$ 24.38\%)} &  \tabincell{c}{76.87\% \\ ($\uparrow$ 54.98\%)} &  \tabincell{c}{73.11\%\\ ($\uparrow$ 43.77\%) }\\ \hline

%     % \hline
% \end{tabular}
% \end{table*}

We explore the effectiveness of \toolName in comparison to the BL method using the default ChatGPT-4o for detecting misconfigurations in serverless applications. 
% The BL method employs an LLM-based approach without incorporating the multi-dimensional constraints that we design.
Table~\ref{tab:RQ2result} presents their results, showing that \toolName is more effective than the BL method. Specifically, \toolName achieves a $precision$ of 72.88\%, $recall$ of 88.18\%, and an $F1$-$score$ of 79.75\%. The BL method achieves a $precision$ of 51.65\%, $recall$ of 65.00\%, and an $F1$-$score$ of 57.55\%. \toolName outperforms the BL method across all metrics, with increases in $precision$ by 21.23 percentage points, $recall$ by 23.18 percentage points, and $F1$-$score$ by 22.20 percentage points.

% shows their specific evaluation metric results. \toolName achieves a precision of XX, recall of XX, and F1-score of XX. In contrast, BLM only reaches a precision of XX\%, recall of XX\%, and F1-score of XX\%. \toolName outperforms BLM by increasing precision by XX percentage points, recall by XX percentage points, and F1-score by XX percentage points, highlighting its superior effectiveness. Different from the BLM method, \toolName uses specific contains from multiple dimensions to make LLMs inferences. The results also illustrate this kind of way is more effective than directly finding misconfigurations in a configuration file using LLMs.

% A key factor contributing to the improved effectiveness of \toolName is its ability to incorporate multi-dimensional constraints for guiding LLM inferences. Unlike the BL method, which directly applies LLMs to detect misconfigurations from configuration files without any optimizations, \toolName leverages guiding thoughts from various dimensions, such as resource type, configuration entries, values, entry dependency, and value dependencies. This enables \toolName to make more accurate and comprehensive inferences, reducing false positives and false negatives.

We investigate the reasons for the low effectiveness of the BL method. Table~\ref{tab:rq2tpfnfptn} shows TP, FN, FP, and TN values obtained by the BL method across all configuration parameters. The results indicate that the BL method has a low TP value of 200, successfully identifying only 64.94\% of the 308 misconfigured parameters. In contrast, \toolName accurately identifies an average of 272 (88.31\%) misconfigured parameters. 
To further explore the root causes of the BL method's low effectiveness, we examine the average number of misconfigured parameters correctly identified across different categories. As presented in Table~\ref{tab:rq2dependency}, the BL method identifies fewer errors than \toolName in each category, including resource types, entries, values, entry dependencies, and value dependencies. Particularly, the BL method only detects 17.95\% of misconfigured entry dependencies, while \toolName detects 97.44\%. We check specific configurations and observe that the BL method struggles to identify configuration entries related to cloud service resources that should co-occur with the event sources defined by serverless functions. For instance, the configuration entry \texttt{RestApiId} under an event source of type ``Api'' should be associated with configuration entries of the ``AWS::Serverless::Api'' resource type.
% We check the detection results of the BL method on 63 misconfigured dependencies from our evaluation dataset, which includes 40 misconfigured entry dependencies and 23 misconfigured value dependencies. Table~\ref{tab:rq2dependency} shows results of detecting misconfigured dependencies across five repetitions. The BL method detects an average of only 17 misconfigured dependencies, whereas \toolName successfully identifies an average of 58. 
Overall, these results indicate that relying solely on the raw capabilities of LLMs, as done in the BL method, is inadequate for the complex task of detecting misconfigurations in serverless applications. A key factor contributing to the improved effectiveness of \toolName is its ability to incorporate multi-dimensional constraints for guiding LLM inferences. These constraints are designed across various dimensions. By integrating them into the analysis, \toolName enhances the decision-making process, resulting in a more effective identification of misconfigurations. 

% This suggests the importance of combining LLM-based reasoning with multi-dimensional constraint information to improve misconfiguration detection effectiveness in serverless applications.

% The evaluation results also indicate that relying solely on the raw capabilities of LLMs, as in the BL method, is insufficient for complex misconfiguration detection tasks of serverless applications. We further observe the effectiveness of the BL method on 63 misconfigured dependencies from our evaluation dataset, which includes 40 misconfigured entry dependencies and 23 misconfigured value dependencies. Table~\ref{tab:rq2dependency} shows results across five repetitions. The BL method detects a mean of only 17 misconfigured dependencies, while \toolName successfully identifies a mean of 58.
% By integrating additional dimensions of analyses, \toolName enhances the decision-making process, leading to more reliable identification of misconfigurations. This highlights the importance of combining LLM-based reasoning with multi-dimensional constraint information for improving the effectiveness of misconfiguration detection in serverless applications.

\begin{table*}[t]
\footnotesize
 \caption{RQ2: The average number of misconfigured parameters correctly identified as misconfigured across different categories.}
 \vspace{-3mm}
    \label{tab:rq2dependency}
    \begin{tabular}{c|c|c|c|c|c}
    \hline
    \textbf{Methods} &  \tabincell{c} {Misconfigured \\resource types (90)} & \tabincell{c} {Misconfigured \\entries (108)} & \tabincell{c} {Misconfigured \\values (48)} & \tabincell{c} {Misconfigured entry \\dependencies (39)} &  \tabincell{c} {Misconfigured value \\ dependencies (23)} \\
    \hline
    BL & 62 (68.89\%) & 83 (76.85\%) & 39 (81.25\%) & 7 (17.95\%) & 12 (52.17\%) \\ \hline
    \toolName & 84 (93.33\%) \checkmark & 93 (86.11\%) \checkmark & 43 (89.58\%) \checkmark & 38 (97.44\%) \checkmark & 19 (82.61\%) \checkmark \\ \hline

\end{tabular}
\vspace{-3mm}
\end{table*}

\finding{\toolName outperforms the BL method across all metrics using the default ChatGPT-4o, with increases in $precision$ by 21.23 percentage points, $recall$ by 23.18 percentage points, and $F1$-$score$ by 22.20 percentage points. This suggests that integrating multi-dimension constraints is beneficial for handling misconfiguration detection in serverless applications.
}

\subsection{RQ3: Impact of Non-determinism on \toolName}

We explore how the non-determinism of LLMs impacts our evaluation results. As detailed in Section~\ref{sec:environment}, each experiment is repeated five times. We analyze their results shown in Table~\ref{tab:rq3} to assess the reliability of our conclusions.
% Table~\ref{tab:rq3} presents the evaluation metrics results of \toolName across five repetitions. 
Results show that while the non-determinism of LLMs can influence evaluation results, its effect is relatively minor. \toolName consistently achieves high effectiveness across different trials. $precision$ ranges from 70.35\% to 76.14\%, $recall$ varies between 84.74\% and 91.88\%, and $F1$-$score$ falls between 76.88\% and 81.21\%. Even the lowest values, i.e., $precision$ at 70.35\%, $recall$ at 84.74\%, $F1$-$score$ at 76.88\%, are still higher than $precision$ (19.06\%), $recall$ (70.78\%), and $F1$-$score$ (30.03\%) of the data-driven approach. Furthermore, the lowest metric values for \toolName remain approximately 20 percentage points higher than the average results (i.e., $precision$ at 51.65\%, $recall$ at 65.00\%, $F1$-$score$ at 57.55\%) of the basic LLM-based method, as shown in Table~\ref{tab:RQ2result}. This suggests that our conclusions regarding \toolName are not affected by the non-determinism of LLMs.

% This RQ explores how the non-determinism of LLMs affects our evaluation results. As described in Section~\ref{sec:environment}, each experiment is repeated five times. We analyze results across these five trials to evaluate the reliability of our conclusion.

% Table~\ref{tab:rq3} presents the results. While the non-determinism of LLMs can influence our evaluation results, the effect is relatively minor. \toolName exhibits consistently high effectiveness in different trials. Specifically, $precision$ keeps from 70.35\% to 76.14\%, $recall$ maintains 84.74\% to 91.88\%, and $F1$-$score$ is ranged from 76.88\% to 81.21\%. Additionally, the lowest $precision$, $recall$, and $F1$-$score$ are still high about 20 percentage points compared to the average result of the BL method shown in Table~\ref{tab:RQ2result}. These results show that our conclusions are less affected by the non-determinism of LLMs.

\begin{table*}[t]
\footnotesize
 \caption{RQ3: Evaluation metrics results of \toolName across five repetitions.}
 \vspace{-3mm}
    \label{tab:rq3}
    \begin{tabular}{l|c|c|c|c|c|c}
    \hline
    \textbf{Metrics} &  Repetition 1 & Repetition 2 & Repetition 3 & Repetition 4 &  Repetition 5 & \textbf{Mean} \\
    \hline
    $precision$ & 71.83\% & 70.78\% & 70.35\% & 75.28\% & 76.14\% & 72.88\% \\ \hline
     $recall$ & 91.88\% & 91.23\% & 84.74\% & 86.04\% & 87.01\% & 88.18\% \\ \hline
     $F1$-$score$ & 80.63\% & 79.72\% & 76.88\% & 80.30\% & 81.21\% & 79.75\% \\ \hline

\end{tabular}
\vspace{-3mm}
\end{table*}

\finding{
Our conclusions are not impacted by the non-determinism of LLMs.
}

\subsection{RQ4: Generalization Capability of \toolName}

\begin{table*}[t]
\footnotesize
 \caption{RQ4: Results about \toolName and BL method using various LLMs.}
 \vspace{-3mm}
    \label{tab:RQ4}
    \begin{tabular}{l|c|c|c||p{2cm}|p{1.4cm}|p{1.4cm}|p{1.4cm}}
    \hline
    \textbf{BL Method} &  $precision$ & $recall$ & $F1$-$score$ & \textbf{Our Approach} &  $precision$ & $recall$ & $F1$-$score$\\
    \hline
    BL (GPT-4o) & 51.65\% & 65.00\%  & 57.55\% & \tabincell{c} {\textbf{\toolName} \\(GPT-4o) (vs BL)} &  \tabincell{c} {72.88\% \\ ($\uparrow$ 21.23\%)} &  \tabincell{c}{88.18\% \\ ($\uparrow$ 23.18\%)} &  \tabincell{c}{79.75\%\\ ($\uparrow$ 22.20\%) }\\ \hline
    BL (Llama) & 48.88\% & 58.38\%  & 53.09\% & \tabincell{c} {\textbf{\toolName}\\ (Llama) (vs BL)} & \tabincell{c} {70.27\% \\ ($\uparrow$ 21.39\%)} &  \tabincell{c}{78.38\% \\ ($\uparrow$ 20.00\%)} &  \tabincell{c}{74.05\%\\ ($\uparrow$ 20.96\%) }\\ \hline
    BL (Gemini) & 44.41\% & 22.86\%  & 30.11\% & \tabincell{c} {\textbf{\toolName} \\(Gemini) (vs BL)} & \tabincell{c} {71.72\% \\ ($\uparrow$ 27.31\%)} &  \tabincell{c}{74.35\% \\ ($\uparrow$ 51.49\%)} &  \tabincell{c}{72.93\%\\ ($\uparrow$ 42.82\%) }\\ \hline

    % \hline
\end{tabular}
\vspace{-3mm}
\end{table*}

% For the implementation of \toolName, we provide flexibility by supporting a variety of LLMs beyond ChatGPT-4o. 
To explore the generalization of \toolName, we use two additional models: the open-source Llama 3.1 (405B) Instruct Turbo model and the proprietary Gemini 1.5 Pro model. \toolName consistently achieves high effectiveness across all metrics, with $precision$, $recall$, and $F1$-$score$ values exceeding 70\%, regardless of the LLM utilized. Table~\ref{tab:RQ4} shows their results. Specifically, with the Llama 3.1 (405B) Instruct Turbo, \toolName achieves a $precision$ of 70.27\%, $recall$ of 78.38\%, and an $F1$-$score$ of 74.05\%. With the Gemini 1.5 Pro model, \toolName yields a $precision$ of 71.72\%, $recall$ of 74.35\%, and an $F1$-$score$ of 72.93\%. Among these, \toolName with ChatGPT-4o offers the highest effectiveness, while \toolName with the Gemini 1.5 Pro model shows comparatively lower metrics but still achieves a high $F1$-$score$ of 72.93\%.

We also evaluate the BL method with different LLMs, shown in Table~\ref{tab:RQ4}. We observe considerable variability. While the BL method achieves $precision$, $recall$, and $F1$-$score$ values approaching or exceeding 50\% when using ChatGPT-4o and Llama 3.1 (405B) Instruct Turbo, its effectiveness drops substantially with the Gemini 1.5 Pro model, where $precision$ is 44.41\%, $recall$ is 22.86\%, and $F1$-$score$ is 30.11\%. This indicates a key limitation of the BL method: its effectiveness is dependent on the specific LLM used. In contrast, \toolName provides the ability to maintain consistent effectiveness across different models, showing its generalization.
% These results of BL methods contrast sharply with the ability of \toolName to maintain consistent effectiveness regardless of the selected model.

We compare the effectiveness differences between \toolName and the BL method when using the same LLM. As discussed in RQ2, \toolName outperforms the BL method with ChatGPT-4o by over 20 percentage points across all evaluation metrics.
% \toolName outperforms the BL method with Chat GPT-4o, higher about 20 percentage points. 
From Table~\ref{tab:RQ4}, when utilizing the Llama 3.1 (405B) Instruct Turbo model, \toolName also achieves improvements of over 20 percentage points across all evaluation metrics compared to the BL method.
% better results compared to the BL methods, with improvements of 21.39 percentage points in $precision$, 20.31 percentage points in $recall$, and 21.09 percentage points in $F1$-$score$. 
% The overall effectiveness difference between \toolName and the BL method with this model is also over 20\% across all three evaluation metrics. 
With the Gemini 1.5 Pro model, \toolName outperforms the BL method with even greater gains, achieving 27.31 percentage points higher in $precision$, 51.49 percentage points higher in $recall$, and 42.82 percentage points higher in $F1$-$score$. The effectiveness gap is especially pronounced with Gemini 1.5 Pro, showing an effectiveness difference of around 50\% in $recall$ and $F1$-$score$, underscoring the effectiveness of our approach.

\finding{
% \toolName does not be influenced by the LLM types, providing consistently high effectiveness, with the precision from XX to XX, recall from XX to XX, and F1-score from XX to XX. For BLM, it is more reliable to use Chat GPT-4o and Llama 3.1 (405B) Instruct Turbo.
\toolName exhibits generalization capability, consistently achieving highly effective results across various LLMs. In contrast, the effectiveness of the BL method varies significantly depending on the chosen LLM. When using the Gemini 1.5 Pro model, \toolName outperforms the BL method by approximately 50 percentage points in both $recall$ and $F1$-$score$.

% With the Gemini 1.5 Pro model, \toolName outperforms BL methods by approximately 50 percentage points in both $recall$ and $F1$-$score$.
}
\section{Threats to Validity}\label{sec:discussion}

\noindent \textbf{Data Leakage Concerns.} One potential risk when using LLMs is data leakage, as these models are trained on vast datasets. Specifically, open-source configuration data may have been exposed to the LLMs utilized in \toolName, raising concerns about memorization of our evaluated error-free configurations available on platforms such as GitHub. However, during our evaluation, we found that the model did not recognize outdated configuration values as correct, suggesting that the error-free configurations we evaluated were not fully present in the LLM’s training data. Note that outdated configuration values were manually corrected before our experimental evaluation.

Our evaluation data also includes both real-world and manually injected misconfigurations. The ground truths for real-world errors are established through an analysis of developer discussions on GitHub to identify root causes. Injected misconfigurations are deliberately introduced into correct configurations through misconfiguration generation rules. These misconfigurations were not exposed to the LLM during training. In addition, the number of configuration files evaluated with errors (58 real-world + 26 injected = 84) significantly exceeded those without errors (26). Thus, the likelihood of our effectiveness results being significantly affected by data leakage is negligible.

We also compare the effectiveness of the BL method without our multi-dimensional constraints. The BL method yields a $precision$ of 51.65\%, a $recall$ of 65.00\%, and an $F1$-$score$ of 57.55\%, indicating low effectiveness. If our evaluation dataset had been exposed to the LLMs, we would expect the BL method to achieve significantly higher results. However, the results do not reflect this, suggesting that our evaluation dataset was not exposed to the LLMs. \toolName incorporates multi-dimensional constraints to detect the same evaluation dataset, resulting in improved metrics: a $precision$ of 72.88\%, a $recall$ of 88.18\%, and an $F1$-$score$ of 79.75\%. These enhancements indicate that our design effectively boosts detection results, rather than being influenced by potential data leakage.

% Thus, the improvements observed can be attributed to the application of our constraints, reinforcing the validity of our evaluation results.

\noindent \textbf{Randomness Concerns.} Randomness can impact evaluation results in two key aspects. The first is the inherent randomness in LLM behavior. To address this, we set a crucial parameter, specifically initializing temperature as 0, to ensure that the model produces consistent outputs for the same input. The second aspect involves randomness in the experimental process. To mitigate this, we conduct five independent experiments for each experimental setup and use the mean results as the final outcome. These strategies minimize the potential impact of randomness on our results.
\section{Related Work}\label{sec:relatedwork}

\subsection{Serverless Computing}

% \noindent \textbf{Serverless Computing.} 

The increasing adoption of serverless computing has attracted widespread interest from the research community, particularly within SE. A broad range of topics has been explored, including literature reviews~\cite{wen2023literature}, evolution and current state~\cite{taibi2020serverlesswhere}, and analyses of serverless application characteristics~\cite{eismann2021state, eismann2020serverless}. Additional research has delved into the challenges faced by developers~\cite{Wen21challenges}, the development of stateful serverless applications~\cite{barcelona2022stateful}, and methods for testing and debugging~\cite{lenarduzzi2020serverless}. For example, a comprehensive literature review~\cite{wen2023literature} was conducted to explore the breadth and depth of serverless computing research. Eismann \textit{et al.}~\cite{eismann2021state} analyzed 89 serverless applications to assess them from multiple dimensions. Wen \textit{et al.}~\cite{Wen21challenges} identified 36 challenges developers face when developing serverless applications, highlighting configuration issues as a prominent concern. Despite these efforts, to the best of our knowledge, no prior work has addressed misconfiguration detection in serverless computing. This paper fills this gap by introducing \toolName.

\subsection{Traditional Misconfiguration Detection}

Existing misconfiguration detection methods can be categorized into two types: white-box and black-box approaches.
White-box approaches~\cite{mehta2020rex, sun2020testing, xu2013not, toman2016staccato, wang2023understanding} generally focus on source code or program analysis to identify misconfigurations within the codebase, relying on manually defined domain-specific rules.
For example, Rex~\cite{mehta2020rex} detected dependency violations between source code and configurations that must be updated together. Ctest~\cite{sun2020testing} identified configuration-induced failures in code affected by configuration changes. SPEX~\cite{xu2013not} employed static program analysis to infer configuration constraints, designing predefined rules from variables in the source code to uncover misconfiguration vulnerabilities. However, these methods are not well-suited for detecting misconfigurations in serverless applications, which rely on YAML-based configuration files rather than source code structures. Serverless-specific misconfigurations, embedded in configuration files, require new approaches that extend beyond traditional white-box techniques.

Black-box approaches~\cite{zhang2014encore, santolucito2016probabilistic, zhou2023drive, santolucito2017synthesizing} are generally data-driven and rely on learning configuration patterns from a dataset of example configurations. 
For instance, EnCore~\cite{zhang2014encore} used numerous configurations to learn and customize rule templates, inferring correlations and detecting misconfigurations in server applications. ConfigC~\cite{santolucito2016probabilistic} analyzed a dataset of correct configurations to build a language model that could detect errors in new configurations. DRIVE~\cite{zhou2023drive} created a Dockerfiles dataset and applied sequential pattern mining to extract frequent patterns, identifying rule violations through heuristic-based reduction and human intervention. 
However, these data-driven methods have inherent limitations: (i) They require a well-curated dataset, but ensuring the completeness and correctness of such datasets is challenging. As a result, configurations not represented in the training data may be missed, while normal configurations might be incorrectly flagged as anomalies due to dataset gaps. (ii) To compensate for dataset issues, these methods incorporate domain-specific knowledge (e.g., customized rule templates), requiring significant manual effort and continuous checking. These limitations hinder the practical application of data-driven approaches.
% for detecting misconfigurations in serverless applications. 
Our results on RQ1 show that such approaches are less effective in our scenario.

\subsection{LLM-based Misconfiguration Detection}

LLM-based approaches offer a promising alternative. A recent arXiv paper presented Ciri~\cite{lian2023configuration}, an LLM-based configuration validator. It demonstrated the potential of LLMs for detecting misconfigurations in systems such as Alluxio, Django, Etcd, and HDFS. However, Ciri depends on an external database containing valid configurations, misconfigurations, related questions, and ground-truth responses. Constructing this database is costly and challenging for various scenarios.
In contrast, \toolName employs zero-shot learning that does not require external datasets, eliminating the need for predefined data. On the other hand, Ciri used a prompt without any constraint, limiting its ability to detect dependencies~\cite{lian2023configuration}. Serverless applications have complex configuration structures and stronger interdependencies, making simple prompt-based methods less effective. Our results on RQ2 and RQ3 show that such a method (i.e., BL method) is less effective in our scenario. Instead, \toolName incorporates carefully designed multi-dimensional constraints without predefined data, providing a more effective detection for serverless application configurations.

\section{Conclusion}\label{sec:conclusion}

Our work opens a promising research direction, showing that LLMs can effectively address configuration issues in cloud applications built on emerging serverless computing.
% Our work provides a potential research direction that LLMs can address configuration problems of cloud applications developed on emerging cloud computing. 
Specifically, we introduced \toolName, the first LLM-based framework specifically designed for detecting misconfigurations in serverless applications. It leveraged advanced prompt engineering and zero-shot learning to effectively identify configuration issues with minimal input effort. \toolName included a prompt generation component that integrates the configuration file to be detected, task description, multi-dimensional constraints, and customized responses. Particularly, the multi-dimensional constraints are tailored to the configuration characteristics of serverless applications, offering context-aware guidance using the Chain of Thought technique. The customized responses focused on both content and format demands, ensuring that the LLM outputs deterministic and clearly explained detection results.

% \toolName included two key modules: the purpose generation module and the answer requirement module, which together construct the final prompt for misconfiguration detection. In the purpose generation module, \toolName integrated the target configuration file, directive questions, and multi-dimensional constraints tailored to the unique characteristics of serverless configurations. These constraints, which are context-aware, provide effective guidance and are implemented using the Chain of Thought technique. The answer requirement module customized both the content and format of the LLM outputs, ensuring that detected misconfigurations are deterministic, clearly explained, and aligned with the expected output structure.

Our evaluation on a curated dataset of 110 configuration files demonstrated that \toolName achieved a precision of 72.88\%, recall of 88.18\%, and F1-score of 79.75\%, surpassing state-of-the-art data-driven methods by 53.82, 17.40, and 49.72 percentage points, respectively. Furthermore, we investigated the generalization capability of \toolName across various LLMs, finding that it consistently maintains high effectiveness across these models.

\bibliographystyle{ACM-Reference-Format}
\bibliography{sample-sigconf}


\begin{thebibliography}{62}

%%% the Web address.

\ifx \showCODEN    \undefined \def \showCODEN     #1{\unskip}     \fi
\ifx \showDOI      \undefined \def \showDOI       #1{#1}\fi
\ifx \showISBNx    \undefined \def \showISBNx     #1{\unskip}     \fi
\ifx \showISBNxiii \undefined \def \showISBNxiii  #1{\unskip}     \fi
\ifx \showISSN     \undefined \def \showISSN      #1{\unskip}     \fi
\ifx \showLCCN     \undefined \def \showLCCN      #1{\unskip}     \fi
\ifx \shownote     \undefined \def \shownote      #1{#1}          \fi
\ifx \showarticletitle \undefined \def \showarticletitle #1{#1}   \fi
\ifx \showURL      \undefined \def \showURL       {\relax}        \fi
% invisible to TeX
\providecommand\bibfield[2]{#2}
\providecommand\bibinfo[2]{#2}
\providecommand\natexlab[1]{#1}
\providecommand\showeprint[2][]{arXiv:#2}

\bibitem[exa(2024)]%
        {examplesam}
 \bibinfo{year}{2024}\natexlab{}.
\newblock \bibinfo{title}{Advantages of building serverless applications on AWS}.
\newblock \bibinfo{howpublished}{\url{https://www.embitel.com/blog/ecommerce-blog/advantages-of-building-serverless-applications-on-aws}}.
\newblock


\bibitem[S3(2024)]%
        {S3}
 \bibinfo{year}{2024}\natexlab{}.
\newblock \bibinfo{title}{Amazon S3}.
\newblock \bibinfo{howpublished}{\url{https://aws.amazon.com/s3/}}.
\newblock


\bibitem[Clo(2024)]%
        {CloudFormation}
 \bibinfo{year}{2024}\natexlab{}.
\newblock \bibinfo{title}{AWS CloudFormation template}.
\newblock \bibinfo{howpublished}{\url{https://docs.aws.amazon.com/AWSCloudFormation/latest/UserGuide/aws-resource-lambda-function.html}}.
\newblock


\bibitem[aws(2024)]%
        {awsresource}
 \bibinfo{year}{2024}\natexlab{}.
\newblock \bibinfo{title}{AWS resource and property types reference}.
\newblock \bibinfo{howpublished}{\url{https://docs.aws.amazon.com/AWSCloudFormation/latest/UserGuide/aws-template-resource-type-ref.html}}.
\newblock


\bibitem[sam(2024)]%
        {sam}
 \bibinfo{year}{2024}\natexlab{}.
\newblock \bibinfo{title}{AWS SAM}.
\newblock \bibinfo{howpublished}{\url{https://aws.amazon.com/cn/serverless/sam}}.
\newblock


\bibitem[ser(2024)]%
        {serverlessresource}
 \bibinfo{year}{2024}\natexlab{}.
\newblock \bibinfo{title}{AWS SAM resources and properties}.
\newblock \bibinfo{howpublished}{\url{https://docs.aws.amazon.com/serverless-application-model/latest/developerguide/sam-specification-resources-and-properties.html}}.
\newblock


\bibitem[SAR(2024)]%
        {SAR}
 \bibinfo{year}{2024}\natexlab{}.
\newblock \bibinfo{title}{AWS Serverless Application Repository}.
\newblock \bibinfo{howpublished}{\url{https://serverlessrepo.aws.amazon.com/applications}}.
\newblock


\bibitem[For(2024)]%
        {FormatVersion}
 \bibinfo{year}{2024}\natexlab{}.
\newblock \bibinfo{title}{AWSTemplateFormatVersion}.
\newblock \bibinfo{howpublished}{\url{https://docs.aws.amazon.com/AWSCloudFormation/latest/UserGuide/format-version-structure.html}}.
\newblock


\bibitem[azu(2024)]%
        {azure}
 \bibinfo{year}{2024}\natexlab{}.
\newblock \bibinfo{title}{Azure Functions}.
\newblock \bibinfo{howpublished}{\url{https://docs.microsoft.com/en-us/azure/azure-functions/}}.
\newblock


\bibitem[CoT(2024)]%
        {CoT}
 \bibinfo{year}{2024}\natexlab{}.
\newblock \bibinfo{title}{Chain-of-Thought Prompting}.
\newblock \bibinfo{howpublished}{\url{https://learnprompting.org/docs/intermediate/chain_of_thought}}.
\newblock


\bibitem[LLM(2024)]%
        {LLMrank}
 \bibinfo{year}{2024}\natexlab{}.
\newblock \bibinfo{title}{Chatbot Arena LLM Leaderboard: Community-driven Evaluation for Best LLM and AI chatbots}.
\newblock \bibinfo{howpublished}{\url{https://lmarena.ai/}}.
\newblock


\bibitem[rea(2024a)]%
        {realexamplemisconfiguration2}
 \bibinfo{year}{2024}\natexlab{a}.
\newblock \bibinfo{title}{DoorDash confirms data breach affected 4.9 million customers, workers and merchants}.
\newblock \bibinfo{howpublished}{\url{https://techcrunch.com/2019/09/26/doordash-data-breach/}}.
\newblock


\bibitem[TES(2024)]%
        {TEST21}
 \bibinfo{year}{2024}\natexlab{}.
\newblock \bibinfo{title}{An example of the configuration file}.
\newblock \bibinfo{howpublished}{\url{https://github.com/aws/serverless-application-model/issues/214}}.
\newblock


\bibitem[goo(2024)]%
        {google}
 \bibinfo{year}{2024}\natexlab{}.
\newblock \bibinfo{title}{Google Cloud Functions}.
\newblock \bibinfo{howpublished}{\url{https://cloud.google.com/functions}}.
\newblock


\bibitem[pop(2024)]%
        {popular1sam}
 \bibinfo{year}{2024}\natexlab{}.
\newblock \bibinfo{title}{Serverless architectures with AWS SAM}.
\newblock \bibinfo{howpublished}{\url{https://medium.com/@christopheradamson253/serverless-architectures-with-aws-sam-serverless-application-model-2b83298fbcbc}}.
\newblock


\bibitem[rea(2024b)]%
        {realexamplemisconfiguration1}
 \bibinfo{year}{2024}\natexlab{b}.
\newblock \bibinfo{title}{Utah COVID-19 testing service}.
\newblock \bibinfo{howpublished}{\url{https://www.comparitech.com/blog/information-security/utah-covid-test-center-leak/}}.
\newblock


\bibitem[Ahmed and Devanbu(2022)]%
        {ahmed2022few}
\bibfield{author}{\bibinfo{person}{Toufique Ahmed} {and} \bibinfo{person}{Premkumar Devanbu}.} \bibinfo{year}{2022}\natexlab{}.
\newblock \showarticletitle{Few-shot training LLMs for project-specific code-summarization}. In \bibinfo{booktitle}{\emph{Proceedings of the 37th IEEE/ACM International Conference on Automated Software Engineering}}. \bibinfo{pages}{1--5}.
\newblock


\bibitem[Ao et~al\mbox{.}(2018)]%
        {ao2018sprocket}
\bibfield{author}{\bibinfo{person}{Lixiang Ao}, \bibinfo{person}{Liz Izhikevich}, \bibinfo{person}{Geoffrey~M Voelker}, {and} \bibinfo{person}{George Porter}.} \bibinfo{year}{2018}\natexlab{}.
\newblock \showarticletitle{Sprocket: A serverless video processing framework}. In \bibinfo{booktitle}{\emph{Proceedings of the ACM Symposium on Cloud Computing}}. \bibinfo{pages}{263--274}.
\newblock


\bibitem[Barcelona-Pons et~al\mbox{.}(2022)]%
        {barcelona2022stateful}
\bibfield{author}{\bibinfo{person}{Daniel Barcelona-Pons}, \bibinfo{person}{Pierre Sutra}, \bibinfo{person}{Marc S{\'a}nchez-Artigas}, \bibinfo{person}{Gerard Par{\'\i}s}, {and} \bibinfo{person}{Pedro Garc{\'\i}a-L{\'o}pez}.} \bibinfo{year}{2022}\natexlab{}.
\newblock \showarticletitle{Stateful serverless computing with crucial}.
\newblock \bibinfo{journal}{\emph{ACM Transactions on Software Engineering and Methodology}} \bibinfo{volume}{31}, \bibinfo{number}{3} (\bibinfo{year}{2022}), \bibinfo{pages}{1--38}.
\newblock


\bibitem[Chen et~al\mbox{.}(2020)]%
        {chen2020understanding}
\bibfield{author}{\bibinfo{person}{Qingrong Chen}, \bibinfo{person}{Teng Wang}, \bibinfo{person}{Owolabi Legunsen}, \bibinfo{person}{Shanshan Li}, {and} \bibinfo{person}{Tianyin Xu}.} \bibinfo{year}{2020}\natexlab{}.
\newblock \showarticletitle{Understanding and discovering software configuration dependencies in cloud and datacenter systems}. In \bibinfo{booktitle}{\emph{Proceedings of the 28th ACM Joint Meeting on European Software Engineering Conference and Symposium on the Foundations of Software Engineering}}. \bibinfo{pages}{362--374}.
\newblock


\bibitem[Chen et~al\mbox{.}(2024)]%
        {chen2024automatic-paper7}
\bibfield{author}{\bibinfo{person}{Yinfang Chen}, \bibinfo{person}{Huaibing Xie}, \bibinfo{person}{Minghua Ma}, \bibinfo{person}{Yu Kang}, \bibinfo{person}{Xin Gao}, \bibinfo{person}{Liu Shi}, \bibinfo{person}{Yunjie Cao}, \bibinfo{person}{Xuedong Gao}, \bibinfo{person}{Hao Fan}, \bibinfo{person}{Ming Wen}, {et~al\mbox{.}}} \bibinfo{year}{2024}\natexlab{}.
\newblock \showarticletitle{Automatic root cause analysis via large language models for cloud incidents}. In \bibinfo{booktitle}{\emph{Proceedings of the 19th European Conference on Computer Systems}}. \bibinfo{pages}{674--688}.
\newblock


\bibitem[Chu et~al\mbox{.}(2023)]%
        {chu2023survey}
\bibfield{author}{\bibinfo{person}{Zheng Chu}, \bibinfo{person}{Jingchang Chen}, \bibinfo{person}{Qianglong Chen}, \bibinfo{person}{Weijiang Yu}, \bibinfo{person}{Tao He}, \bibinfo{person}{Haotian Wang}, \bibinfo{person}{Weihua Peng}, \bibinfo{person}{Ming Liu}, \bibinfo{person}{Bing Qin}, {and} \bibinfo{person}{Ting Liu}.} \bibinfo{year}{2023}\natexlab{}.
\newblock \showarticletitle{A survey of chain of thought reasoning: Advances, frontiers and future}.
\newblock \bibinfo{journal}{\emph{arXiv preprint arXiv:2309.15402}} (\bibinfo{year}{2023}).
\newblock


\bibitem[Cohen(1960)]%
        {cohen1960coefficient}
\bibfield{author}{\bibinfo{person}{Jacob Cohen}.} \bibinfo{year}{1960}\natexlab{}.
\newblock \showarticletitle{A coefficient of agreement for nominal scales}.
\newblock \bibinfo{journal}{\emph{Educational and psychological measurement}} \bibinfo{volume}{20}, \bibinfo{number}{1} (\bibinfo{year}{1960}), \bibinfo{pages}{37--46}.
\newblock


\bibitem[Dong et~al\mbox{.}(2024)]%
        {dong2024self}
\bibfield{author}{\bibinfo{person}{Yihong Dong}, \bibinfo{person}{Xue Jiang}, \bibinfo{person}{Zhi Jin}, {and} \bibinfo{person}{Ge Li}.} \bibinfo{year}{2024}\natexlab{}.
\newblock \showarticletitle{Self-collaboration code generation via ChatGPT}.
\newblock \bibinfo{journal}{\emph{ACM Transactions on Software Engineering and Methodology}} \bibinfo{volume}{33}, \bibinfo{number}{7} (\bibinfo{year}{2024}), \bibinfo{pages}{1--38}.
\newblock


\bibitem[Eismann et~al\mbox{.}(2021)]%
        {eismann2021state}
\bibfield{author}{\bibinfo{person}{Simon Eismann}, \bibinfo{person}{Joel Scheuner}, \bibinfo{person}{Erwin~Van Eyk}, \bibinfo{person}{Maximilian Schwinger}, \bibinfo{person}{Johannes Grohmann}, \bibinfo{person}{Nikolas Herbst}, \bibinfo{person}{Cristina Abad}, {and} \bibinfo{person}{Alexandru Iosup}.} \bibinfo{year}{2021}\natexlab{}.
\newblock \showarticletitle{The state of serverless applications: collection, characterization, and community consensus}.
\newblock \bibinfo{journal}{\emph{IEEE Transactions on Software Engineering}} (\bibinfo{year}{2021}).
\newblock


\bibitem[Eismann et~al\mbox{.}(2020)]%
        {eismann2020serverless}
\bibfield{author}{\bibinfo{person}{Simon Eismann}, \bibinfo{person}{Joel Scheuner}, \bibinfo{person}{Erwin Van~Eyk}, \bibinfo{person}{Maximilian Schwinger}, \bibinfo{person}{Johannes Grohmann}, \bibinfo{person}{Nikolas Herbst}, \bibinfo{person}{Cristina~L Abad}, {and} \bibinfo{person}{Alexandru Iosup}.} \bibinfo{year}{2020}\natexlab{}.
\newblock \showarticletitle{Serverless applications: Why, when, and how?}
\newblock \bibinfo{journal}{\emph{IEEE Software}} \bibinfo{volume}{38}, \bibinfo{number}{1} (\bibinfo{year}{2020}), \bibinfo{pages}{32--39}.
\newblock


\bibitem[Fan et~al\mbox{.}(2023)]%
        {fan2023automated}
\bibfield{author}{\bibinfo{person}{Zhiyu Fan}, \bibinfo{person}{Xiang Gao}, \bibinfo{person}{Martin Mirchev}, \bibinfo{person}{Abhik Roychoudhury}, {and} \bibinfo{person}{Shin~Hwei Tan}.} \bibinfo{year}{2023}\natexlab{}.
\newblock \showarticletitle{Automated repair of programs from large language models}. In \bibinfo{booktitle}{\emph{Proceedings of the 2023 IEEE/ACM 45th International Conference on Software Engineering}}. IEEE, \bibinfo{pages}{1469--1481}.
\newblock


\bibitem[Hadadi et~al\mbox{.}(2024)]%
        {hadadi2024anomaly}
\bibfield{author}{\bibinfo{person}{Fatemeh Hadadi}, \bibinfo{person}{Qinghua Xu}, \bibinfo{person}{Domenico Bianculli}, {and} \bibinfo{person}{Lionel Briand}.} \bibinfo{year}{2024}\natexlab{}.
\newblock \showarticletitle{Anomaly detection on unstable logs with GPT models}.
\newblock \bibinfo{journal}{\emph{arXiv preprint arXiv:2406.07467}} (\bibinfo{year}{2024}).
\newblock


\bibitem[Han et~al\mbox{.}(2000)]%
        {han2000mining}
\bibfield{author}{\bibinfo{person}{Jiawei Han}, \bibinfo{person}{Jian Pei}, {and} \bibinfo{person}{Yiwen Yin}.} \bibinfo{year}{2000}\natexlab{}.
\newblock \showarticletitle{Mining frequent patterns without candidate generation}.
\newblock \bibinfo{journal}{\emph{ACM SIGMOD Record}} \bibinfo{volume}{29}, \bibinfo{number}{2} (\bibinfo{year}{2000}), \bibinfo{pages}{1--12}.
\newblock


\bibitem[Hassan et~al\mbox{.}(2021)]%
        {hassan2021survey}
\bibfield{author}{\bibinfo{person}{Hassan~B Hassan}, \bibinfo{person}{Saman~A Barakat}, {and} \bibinfo{person}{Qusay~I Sarhan}.} \bibinfo{year}{2021}\natexlab{}.
\newblock \showarticletitle{Survey on serverless computing}.
\newblock \bibinfo{journal}{\emph{Journal of Cloud Computing}} \bibinfo{volume}{10}, \bibinfo{number}{1} (\bibinfo{year}{2021}), \bibinfo{pages}{1--29}.
\newblock


\bibitem[Imran et~al\mbox{.}(2024)]%
        {imran2024uncovering}
\bibfield{author}{\bibinfo{person}{Mia~Mohammad Imran}, \bibinfo{person}{Preetha Chatterjee}, {and} \bibinfo{person}{Kostadin Damevski}.} \bibinfo{year}{2024}\natexlab{}.
\newblock \showarticletitle{Uncovering the causes of emotions in software developer communication using zero-shot llms}. In \bibinfo{booktitle}{\emph{Proceedings of the IEEE/ACM 46th International Conference on Software Engineering}}. \bibinfo{pages}{1--13}.
\newblock


\bibitem[Jonas et~al\mbox{.}(2019)]%
        {JonasCoRR2019}
\bibfield{author}{\bibinfo{person}{Eric Jonas}, \bibinfo{person}{Johann Schleier-Smith}, \bibinfo{person}{Vikram Sreekanti}, \bibinfo{person}{Chia-Che Tsai}, \bibinfo{person}{Anurag Khandelwal}, \bibinfo{person}{Qifan Pu}, \bibinfo{person}{Vaishaal Shankar}, \bibinfo{person}{Joao Carreira}, \bibinfo{person}{Karl Krauth}, \bibinfo{person}{Neeraja Yadwadkar}, \bibinfo{person}{Joseph~E. Gonzalez}, \bibinfo{person}{Raluca~Ada Popa}, \bibinfo{person}{Ion Stoica}, {and} \bibinfo{person}{David~A. Patterson}.} \bibinfo{year}{2019}\natexlab{}.
\newblock \showarticletitle{Cloud programming simplified: A Berkeley view on serverless computing}.
\newblock \bibinfo{journal}{\emph{arXiv preprint arXiv:1902.03383}} (\bibinfo{year}{2019}).
\newblock


\bibitem[Kavaler et~al\mbox{.}(2017)]%
        {kavaler2017perceived}
\bibfield{author}{\bibinfo{person}{David Kavaler}, \bibinfo{person}{Sasha Sirovica}, \bibinfo{person}{Vincent Hellendoorn}, \bibinfo{person}{Raul Aranovich}, {and} \bibinfo{person}{Vladimir Filkov}.} \bibinfo{year}{2017}\natexlab{}.
\newblock \showarticletitle{Perceived language complexity in GitHub issue discussions and their effect on issue resolution}. In \bibinfo{booktitle}{\emph{Proceedings of the 2017 32nd IEEE/ACM International Conference on Automated Software Engineering}}. IEEE, \bibinfo{pages}{72--83}.
\newblock


\bibitem[Kourtzanidis et~al\mbox{.}(2020)]%
        {kourtzanidis2020reposkillminer}
\bibfield{author}{\bibinfo{person}{Stratos Kourtzanidis}, \bibinfo{person}{Alexander Chatzigeorgiou}, {and} \bibinfo{person}{Apostolos Ampatzoglou}.} \bibinfo{year}{2020}\natexlab{}.
\newblock \showarticletitle{RepoSkillMiner: Identifying software expertise from GitHub repositories using natural language processing}. In \bibinfo{booktitle}{\emph{Proceedings of the 35th IEEE/ACM International Conference on Automated Software Engineering}}. \bibinfo{pages}{1353--1357}.
\newblock


\bibitem[Landis and Koch(1977)]%
        {landis1977measurement}
\bibfield{author}{\bibinfo{person}{J~Richard Landis} {and} \bibinfo{person}{Gary~G Koch}.} \bibinfo{year}{1977}\natexlab{}.
\newblock \showarticletitle{The measurement of observer agreement for categorical data}.
\newblock \bibinfo{journal}{\emph{Biometrics}} \bibinfo{volume}{33}, \bibinfo{number}{1} (\bibinfo{year}{1977}), \bibinfo{pages}{159--174}.
\newblock


\bibitem[Lenarduzzi and Panichella(2020)]%
        {lenarduzzi2020serverless}
\bibfield{author}{\bibinfo{person}{Valentina Lenarduzzi} {and} \bibinfo{person}{Annibale Panichella}.} \bibinfo{year}{2020}\natexlab{}.
\newblock \showarticletitle{Serverless testing: Tool vendors' and experts' points of view}.
\newblock \bibinfo{journal}{\emph{IEEE Software}} \bibinfo{volume}{38}, \bibinfo{number}{1} (\bibinfo{year}{2020}), \bibinfo{pages}{54--60}.
\newblock


\bibitem[Li et~al\mbox{.}(2023)]%
        {li2023structured}
\bibfield{author}{\bibinfo{person}{Jia Li}, \bibinfo{person}{Ge Li}, \bibinfo{person}{Yongmin Li}, {and} \bibinfo{person}{Zhi Jin}.} \bibinfo{year}{2023}\natexlab{}.
\newblock \showarticletitle{Structured chain-of-thought prompting for code generation}.
\newblock \bibinfo{journal}{\emph{ACM Transactions on Software Engineering and Methodology}} (\bibinfo{year}{2023}).
\newblock


\bibitem[Li et~al\mbox{.}(2018)]%
        {li2018confvd}
\bibfield{author}{\bibinfo{person}{Shanshan Li}, \bibinfo{person}{Wang Li}, \bibinfo{person}{Xiangke Liao}, \bibinfo{person}{Shaoliang Peng}, \bibinfo{person}{Shulin Zhou}, \bibinfo{person}{Zhouyang Jia}, {and} \bibinfo{person}{Teng Wang}.} \bibinfo{year}{2018}\natexlab{}.
\newblock \showarticletitle{Confvd: System reactions analysis and evaluation through misconfiguration injection}.
\newblock \bibinfo{journal}{\emph{IEEE Transactions on Reliability}} \bibinfo{volume}{67}, \bibinfo{number}{4} (\bibinfo{year}{2018}), \bibinfo{pages}{1393--1405}.
\newblock


\bibitem[Li et~al\mbox{.}(2021)]%
        {li2021challenges}
\bibfield{author}{\bibinfo{person}{Wang Li}, \bibinfo{person}{Zhouyang Jia}, \bibinfo{person}{Shanshan Li}, \bibinfo{person}{Yuanliang Zhang}, \bibinfo{person}{Teng Wang}, \bibinfo{person}{Erci Xu}, \bibinfo{person}{Ji Wang}, {and} \bibinfo{person}{Xiangke Liao}.} \bibinfo{year}{2021}\natexlab{}.
\newblock \showarticletitle{Challenges and opportunities: An in-depth empirical study on configuration error injection testing}. In \bibinfo{booktitle}{\emph{Proceedings of the 30th ACM SIGSOFT International Symposium on Software Testing and Analysis}}. \bibinfo{pages}{478--490}.
\newblock


\bibitem[Li et~al\mbox{.}(2022)]%
        {LiGCCHG22}
\bibfield{author}{\bibinfo{person}{Zijun Li}, \bibinfo{person}{Linsong Guo}, \bibinfo{person}{Jiagan Cheng}, \bibinfo{person}{Quan Chen}, \bibinfo{person}{Bingsheng He}, {and} \bibinfo{person}{Minyi Guo}.} \bibinfo{year}{2022}\natexlab{}.
\newblock \showarticletitle{The serverless computing survey: {A} technical primer for design architecture}.
\newblock \bibinfo{journal}{\emph{Comput. Surveys}} \bibinfo{volume}{54}, \bibinfo{number}{10s} (\bibinfo{year}{2022}), \bibinfo{pages}{220:1--220:34}.
\newblock


\bibitem[Lian et~al\mbox{.}(2023)]%
        {lian2023configuration}
\bibfield{author}{\bibinfo{person}{Xinyu Lian}, \bibinfo{person}{Yinfang Chen}, \bibinfo{person}{Runxiang Cheng}, \bibinfo{person}{Jie Huang}, \bibinfo{person}{Parth Thakkar}, {and} \bibinfo{person}{Tianyin Xu}.} \bibinfo{year}{2023}\natexlab{}.
\newblock \showarticletitle{Configuration validation with large language models}.
\newblock \bibinfo{journal}{\emph{arXiv preprint arXiv:2310.09690}} (\bibinfo{year}{2023}).
\newblock


\bibitem[Mehta et~al\mbox{.}(2020)]%
        {mehta2020rex}
\bibfield{author}{\bibinfo{person}{Sonu Mehta}, \bibinfo{person}{Ranjita Bhagwan}, \bibinfo{person}{Rahul Kumar}, \bibinfo{person}{Chetan Bansal}, \bibinfo{person}{Chandra Maddila}, \bibinfo{person}{Balasubramanyan Ashok}, \bibinfo{person}{Sumit Asthana}, \bibinfo{person}{Christian Bird}, {and} \bibinfo{person}{Aditya Kumar}.} \bibinfo{year}{2020}\natexlab{}.
\newblock \showarticletitle{Rex: Preventing bugs and misconfiguration in large services using correlated change analysis}. In \bibinfo{booktitle}{\emph{Proceedings of the 17th USENIX Symposium on Networked Systems Design and Implementation}}. \bibinfo{pages}{435--448}.
\newblock


\bibitem[Santolucito et~al\mbox{.}(2017)]%
        {santolucito2017synthesizing}
\bibfield{author}{\bibinfo{person}{Mark Santolucito}, \bibinfo{person}{Ennan Zhai}, \bibinfo{person}{Rahul Dhodapkar}, \bibinfo{person}{Aaron Shim}, {and} \bibinfo{person}{Ruzica Piskac}.} \bibinfo{year}{2017}\natexlab{}.
\newblock \showarticletitle{Synthesizing configuration file specifications with association rule learning}.
\newblock \bibinfo{journal}{\emph{Proceedings of the ACM on Programming Languages}} \bibinfo{volume}{1}, \bibinfo{number}{OOPSLA} (\bibinfo{year}{2017}), \bibinfo{pages}{1--20}.
\newblock


\bibitem[Santolucito et~al\mbox{.}(2016)]%
        {santolucito2016probabilistic}
\bibfield{author}{\bibinfo{person}{Mark Santolucito}, \bibinfo{person}{Ennan Zhai}, {and} \bibinfo{person}{Ruzica Piskac}.} \bibinfo{year}{2016}\natexlab{}.
\newblock \showarticletitle{Probabilistic automated language learning for configuration files}. In \bibinfo{booktitle}{\emph{Proceedings of the Computer Aided Verification: 28th International Conference}}. Springer, \bibinfo{pages}{80--87}.
\newblock


\bibitem[Shankar et~al\mbox{.}(2020)]%
        {shankar2020serverless}
\bibfield{author}{\bibinfo{person}{Vaishaal Shankar}, \bibinfo{person}{Karl Krauth}, \bibinfo{person}{Kailas Vodrahalli}, \bibinfo{person}{Qifan Pu}, \bibinfo{person}{Benjamin Recht}, \bibinfo{person}{Ion Stoica}, \bibinfo{person}{Jonathan Ragan-Kelley}, \bibinfo{person}{Eric Jonas}, {and} \bibinfo{person}{Shivaram Venkataraman}.} \bibinfo{year}{2020}\natexlab{}.
\newblock \showarticletitle{Serverless linear algebra}. In \bibinfo{booktitle}{\emph{Proceedings of the 11th ACM Symposium on Cloud Computing}}. \bibinfo{pages}{281--295}.
\newblock


\bibitem[Sun et~al\mbox{.}(2020)]%
        {sun2020testing}
\bibfield{author}{\bibinfo{person}{Xudong Sun}, \bibinfo{person}{Runxiang Cheng}, \bibinfo{person}{Jianyan Chen}, \bibinfo{person}{Elaine Ang}, \bibinfo{person}{Owolabi Legunsen}, {and} \bibinfo{person}{Tianyin Xu}.} \bibinfo{year}{2020}\natexlab{}.
\newblock \showarticletitle{Testing configuration changes in context to prevent production failures}. In \bibinfo{booktitle}{\emph{Proceedings of the 14th USENIX Symposium on Operating Systems Design and Implementation}}. \bibinfo{pages}{735--751}.
\newblock


\bibitem[Taibi et~al\mbox{.}(2020)]%
        {taibi2020serverlesswhere}
\bibfield{author}{\bibinfo{person}{Davide Taibi}, \bibinfo{person}{Josef Spillner}, {and} \bibinfo{person}{Konrad Wawruch}.} \bibinfo{year}{2020}\natexlab{}.
\newblock \showarticletitle{Serverless computing-where are we now, and where are we heading?}
\newblock \bibinfo{journal}{\emph{IEEE software}} \bibinfo{volume}{38}, \bibinfo{number}{1} (\bibinfo{year}{2020}), \bibinfo{pages}{25--31}.
\newblock


\bibitem[Toman and Grossman(2016)]%
        {toman2016staccato}
\bibfield{author}{\bibinfo{person}{John Toman} {and} \bibinfo{person}{Dan Grossman}.} \bibinfo{year}{2016}\natexlab{}.
\newblock \showarticletitle{Staccato: A bug finder for dynamic configuration updates}. In \bibinfo{booktitle}{\emph{Proceedings of the 30th European Conference on Object-Oriented Programming}}. Schloss Dagstuhl-Leibniz-Zentrum fuer Informatik.
\newblock


\bibitem[Wang et~al\mbox{.}(2023)]%
        {wang2023understanding}
\bibfield{author}{\bibinfo{person}{Teng Wang}, \bibinfo{person}{Zhouyang Jia}, \bibinfo{person}{Shanshan Li}, \bibinfo{person}{Si Zheng}, \bibinfo{person}{Yue Yu}, \bibinfo{person}{Erci Xu}, \bibinfo{person}{Shaoliang Peng}, {and} \bibinfo{person}{Xiangke Liao}.} \bibinfo{year}{2023}\natexlab{}.
\newblock \showarticletitle{Understanding and detecting on-the-fly configuration bugs}. In \bibinfo{booktitle}{\emph{Proceedings of the 45th International Conference on Software Engineering}}.
\newblock


\bibitem[Wen et~al\mbox{.}(2023)]%
        {wen2023literature}
\bibfield{author}{\bibinfo{person}{Jinfeng Wen}, \bibinfo{person}{Zhenpeng Chen}, \bibinfo{person}{Xin Jin}, {and} \bibinfo{person}{Xuanzhe Liu}.} \bibinfo{year}{2023}\natexlab{}.
\newblock \showarticletitle{Rise of the planet of serverless computing: a systematic review}.
\newblock \bibinfo{journal}{\emph{ACM Transactions on Software Engineering and Methodology}} \bibinfo{volume}{32}, \bibinfo{number}{5} (\bibinfo{year}{2023}), \bibinfo{pages}{1--61}.
\newblock


\bibitem[Wen et~al\mbox{.}(2021)]%
        {Wen21challenges}
\bibfield{author}{\bibinfo{person}{Jinfeng Wen}, \bibinfo{person}{Zhenpeng Chen}, \bibinfo{person}{Yi Liu}, \bibinfo{person}{Yiling Lou}, \bibinfo{person}{Yun Ma}, \bibinfo{person}{Gang Huang}, \bibinfo{person}{Xin Jin}, {and} \bibinfo{person}{Xuanzhe Liu}.} \bibinfo{year}{2021}\natexlab{}.
\newblock \showarticletitle{An empirical study on challenges of application development in serverless computing}. In \bibinfo{booktitle}{\emph{Proceedings of the 28th ACM Joint Meeting on European Software Engineering Conference and Symposium on the Foundations of Software Engineering}}. \bibinfo{pages}{416--428}.
\newblock


\bibitem[Xie et~al\mbox{.}(2023)]%
        {xie2023chatunitest}
\bibfield{author}{\bibinfo{person}{Zhuokui Xie}, \bibinfo{person}{Yinghao Chen}, \bibinfo{person}{Chen Zhi}, \bibinfo{person}{Shuiguang Deng}, {and} \bibinfo{person}{Jianwei Yin}.} \bibinfo{year}{2023}\natexlab{}.
\newblock \showarticletitle{ChatUniTest: a ChatGPT-based automated unit test generation tool}.
\newblock \bibinfo{journal}{\emph{arXiv preprint arXiv:2305.04764}} (\bibinfo{year}{2023}).
\newblock


\bibitem[Xu et~al\mbox{.}(2024)]%
        {xu2024divlog}
\bibfield{author}{\bibinfo{person}{Junjielong Xu}, \bibinfo{person}{Ruichun Yang}, \bibinfo{person}{Yintong Huo}, \bibinfo{person}{Chengyu Zhang}, {and} \bibinfo{person}{Pinjia He}.} \bibinfo{year}{2024}\natexlab{}.
\newblock \showarticletitle{DivLog: Log parsing with prompt enhanced in-context learning}. In \bibinfo{booktitle}{\emph{Proceedings of the IEEE/ACM 46th International Conference on Software Engineering}}. \bibinfo{pages}{1--12}.
\newblock


\bibitem[Xu et~al\mbox{.}(2013)]%
        {xu2013not}
\bibfield{author}{\bibinfo{person}{Tianyin Xu}, \bibinfo{person}{Jiaqi Zhang}, \bibinfo{person}{Peng Huang}, \bibinfo{person}{Jing Zheng}, \bibinfo{person}{Tianwei Sheng}, \bibinfo{person}{Ding Yuan}, \bibinfo{person}{Yuanyuan Zhou}, {and} \bibinfo{person}{Shankar Pasupathy}.} \bibinfo{year}{2013}\natexlab{}.
\newblock \showarticletitle{Do not blame users for misconfigurations}. In \bibinfo{booktitle}{\emph{Proceedings of the 24th ACM Symposium on Operating Systems Principles}}. \bibinfo{pages}{244--259}.
\newblock


\bibitem[Xu and Zhou(2015)]%
        {xu2015systems}
\bibfield{author}{\bibinfo{person}{Tianyin Xu} {and} \bibinfo{person}{Yuanyuan Zhou}.} \bibinfo{year}{2015}\natexlab{}.
\newblock \showarticletitle{Systems approaches to tackling configuration errors: A survey}.
\newblock \bibinfo{journal}{\emph{Comput. Surveys}} \bibinfo{volume}{47}, \bibinfo{number}{4} (\bibinfo{year}{2015}), \bibinfo{pages}{1--41}.
\newblock


\bibitem[Xu et~al\mbox{.}(2009)]%
        {xu2009detecting}
\bibfield{author}{\bibinfo{person}{Wei Xu}, \bibinfo{person}{Ling Huang}, \bibinfo{person}{Armando Fox}, \bibinfo{person}{David Patterson}, {and} \bibinfo{person}{Michael~I Jordan}.} \bibinfo{year}{2009}\natexlab{}.
\newblock \showarticletitle{Detecting large-scale system problems by mining console logs}. In \bibinfo{booktitle}{\emph{Proceedings of the ACM SIGOPS 22nd symposium on Operating systems principles}}. \bibinfo{pages}{117--132}.
\newblock


\bibitem[Yin et~al\mbox{.}(2024)]%
        {yin2024multitask}
\bibfield{author}{\bibinfo{person}{Xin Yin}, \bibinfo{person}{Chao Ni}, {and} \bibinfo{person}{Shaohua Wang}.} \bibinfo{year}{2024}\natexlab{}.
\newblock \showarticletitle{Multitask-based evaluation of open-source llm on software vulnerability}.
\newblock \bibinfo{journal}{\emph{IEEE Transactions on Software Engineering}} (\bibinfo{year}{2024}).
\newblock


\bibitem[Yu et~al\mbox{.}(2021)]%
        {yu2021gillis}
\bibfield{author}{\bibinfo{person}{Minchen Yu}, \bibinfo{person}{Zhifeng Jiang}, \bibinfo{person}{Hok~Chun Ng}, \bibinfo{person}{Wei Wang}, \bibinfo{person}{Ruichuan Chen}, {and} \bibinfo{person}{Bo Li}.} \bibinfo{year}{2021}\natexlab{}.
\newblock \showarticletitle{Gillis: Serving large neural networks in serverless functions with automatic model partitioning}. In \bibinfo{booktitle}{\emph{Proceedings of the 2021 IEEE 41st International Conference on Distributed Computing Systems}}. IEEE, \bibinfo{pages}{138--148}.
\newblock


\bibitem[Yuan et~al\mbox{.}(2011)]%
        {yuan2011context}
\bibfield{author}{\bibinfo{person}{Ding Yuan}, \bibinfo{person}{Yinglian Xie}, \bibinfo{person}{Rina Panigrahy}, \bibinfo{person}{Junfeng Yang}, \bibinfo{person}{Chad Verbowski}, {and} \bibinfo{person}{Arunvijay Kumar}.} \bibinfo{year}{2011}\natexlab{}.
\newblock \showarticletitle{Context-based online configuration-error detection}. In \bibinfo{booktitle}{\emph{Proceedings of the 2011 USENIX Conference on USENIX Annual Technical Conference}}. \bibinfo{pages}{28--28}.
\newblock


\bibitem[Yuan et~al\mbox{.}(2024)]%
        {yuan2024evaluating}
\bibfield{author}{\bibinfo{person}{Zhiqiang Yuan}, \bibinfo{person}{Mingwei Liu}, \bibinfo{person}{Shiji Ding}, \bibinfo{person}{Kaixin Wang}, \bibinfo{person}{Yixuan Chen}, \bibinfo{person}{Xin Peng}, {and} \bibinfo{person}{Yiling Lou}.} \bibinfo{year}{2024}\natexlab{}.
\newblock \showarticletitle{Evaluating and improving ChatGPT for unit test generation}.
\newblock \bibinfo{journal}{\emph{Proceedings of the ACM on Software Engineering}} \bibinfo{volume}{1}, \bibinfo{number}{FSE} (\bibinfo{year}{2024}), \bibinfo{pages}{1703--1726}.
\newblock


\bibitem[Zhang et~al\mbox{.}(2014)]%
        {zhang2014encore}
\bibfield{author}{\bibinfo{person}{Jiaqi Zhang}, \bibinfo{person}{Lakshminarayanan Renganarayana}, \bibinfo{person}{Xiaolan Zhang}, \bibinfo{person}{Niyu Ge}, \bibinfo{person}{Vasanth Bala}, \bibinfo{person}{Tianyin Xu}, {and} \bibinfo{person}{Yuanyuan Zhou}.} \bibinfo{year}{2014}\natexlab{}.
\newblock \showarticletitle{EnCore: Exploiting system environment and correlation information for misconfiguration detection}. In \bibinfo{booktitle}{\emph{Proceedings of the 19th International Conference on Architectural Support for Programming Languages and Operating Systems}}. \bibinfo{pages}{687--700}.
\newblock


\bibitem[Zhou et~al\mbox{.}(2023)]%
        {zhou2023drive}
\bibfield{author}{\bibinfo{person}{Yu Zhou}, \bibinfo{person}{Weilin Zhan}, \bibinfo{person}{Zi Li}, \bibinfo{person}{Tingting Han}, \bibinfo{person}{Taolue Chen}, {and} \bibinfo{person}{Harald Gall}.} \bibinfo{year}{2023}\natexlab{}.
\newblock \showarticletitle{DRIVE: Dockerfile rule mining and violation detection}.
\newblock \bibinfo{journal}{\emph{ACM Transactions on Software Engineering and Methodology}} \bibinfo{volume}{33}, \bibinfo{number}{2} (\bibinfo{year}{2023}), \bibinfo{pages}{1--23}.
\newblock


\end{thebibliography}


\end{document}